\documentclass[%
 reprint,
superscriptaddress,
 amsmath,amssymb,
 aps,
]{revtex4-2}

\usepackage{graphicx}
\usepackage{dcolumn}
\usepackage{bm}
\usepackage{xcolor}

\begin{document}

\preprint{APS/123-QED}

\title{Learning the tensor network model of a quantum state using a few single-qubit measurements}

\author{Sergei Kuzmin}
\email{s.kuzmin@rqc.ru}
\affiliation{Quantum Technologies Centre, Lomonosov Moscow State University, Russia, Moscow, 119991, Leninskie Gory 1 building 35}
\affiliation{Russian Quantum Center, Russia, Moscow, 121205, Bol'shoy bul'var 30 building 1}
\author{Varvara Mikhailova}%
\affiliation{Quantum Technologies Centre, Lomonosov Moscow State University, Russia, Moscow, 119991, Leninskie Gory 1 building 35}
\affiliation{Russian Quantum Center, Russia, Moscow, 121205, Bol'shoy bul'var 30 building 1}
\author{Ivan Dyakonov}%
\email{iv.dyakonov@quantum.msu.ru}
\affiliation{Quantum Technologies Centre, Lomonosov Moscow State University, Russia, Moscow, 119991, Leninskie Gory 1 building 35}
\affiliation{Russian Quantum Center, Russia, Moscow, 121205, Bol'shoy bul'var 30 building 1}
\author{Stanislav Straupe}%
\affiliation{Quantum Technologies Centre, Lomonosov Moscow State University, Russia, Moscow, 119991, Leninskie Gory 1 building 35}
\affiliation{Russian Quantum Center, Russia, Moscow, 121205, Bol'shoy bul'var 30 building 1}

\date{\today}

\begin{abstract}
The constantly increasing dimensionality of artificial quantum systems demands for highly efficient methods for their characterization and benchmarking. Conventional quantum tomography fails for larger systems due to the exponential growth of the required number of measurements. The conceptual solution for this dimensionality curse relies on a simple idea -- a complete description of a quantum state is excessive and can be discarded in favor of experimentally accessible information about the system. The probably approximately correct (PAC) learning theory has been recently successfully applied to a problem of building accurate predictors for the measurement outcomes using a dataset which scales only linearly with the number of qubits. Here we present a constructive and numerically efficient protocol which learns a tensor network model of an unknown quantum system. We discuss the limitations and the scalability of the proposed method.
\end{abstract}

\maketitle


\section{\label{sec:Introduction}Introduction}

The field of quantum computing has experienced unprecedented growth in the last decade. The main reason is the emergence of experimental prototypes of quantum processors with dozens of well-controlled qubits. These devices have not yet reached the error-correction threshold and belong to the so-called NISQ (Noisy Intermediate-Scale Quantum) class of processors, however their complexity reaches the borderline of the classical computer simulation power. The advent of the NISQ technology culminated in the demonstration of quantum computational advantage by the Google AI team using a superconducting processor with 53 qubits \cite{Quantum_Supremacy_Google}. Other experimental platforms, such as trapped ions \cite{Ions_Trapped_Wright2019}, neutral atoms \cite{Neutral_Atoms_Ebadi2021}, and photons \cite{Photons_Intro}, are also becoming increasingly competitive. Further development and scaling of these devices requires new tools to describe their features and benchmark their performance.

The paramount task is to infer the features of a multi-qubit quantum system \cite{10.1063/1.1494475, PhysRevLett.98.020403}. The complete reconstruction of the density operator $\hat{\rho}$ of $N$ qubits requires an exponential number of $\sim 4^N$ repeated measurements and classical processing \cite{Haah_2017}. Therefore, full quantum state tomography (QST) can not be considered a practical tool for mid- and large-scale systems. However, there are several approaches designed to partially solve this problem in some cases, in particular, using the neural network approach \cite{torlai2018neural, palmieri2020experimental, quek2021adaptive, jia2019quantum}, as well as using the tensor networks approach \cite{ran2020tensor, Torlai2023} it is possible to achieve a reduction in the volume of measurements.

In practice a quantum state description is always accompanied by a quality estimator -- fidelity or other measure of interest. However, information contained in the full quantum state is usually redundant for simple benchmarks. The seminal work \cite{Aaronson2007} applied the PAC learning theory \cite{valiant1984theory} to answer a simple question -- whether the success probability of the two-outcome POVM measurement can be predicted efficiently? It turned out that this problem is positively resolved employing only $O(N)$ measurements. This result sparked the interest and a series of works \cite{PRXQuantum.2.030348, Hadfield2022, PhysRevLett.127.110504, levy2021classical, PhysRevLett.126.050501, PRXQuantum.4.010303, Elben_2022} studied this approach which is now generally referred to as shadow tomography. The theoretical studies were followed by experimental applications using an optical processor \cite{doi:10.1126/sciadv.aau1946}. Further research unveiled that the certain features such as fidelity and average values of low-rank observables can be estimated using a constant number of state copies independent of $N$ \cite{Huang2020}. This result is proven to be optimal from the information-theoretic viewpoint, however it is experimentally challenging due to a very specific measurement set (randomized multi-qubit Clifford measurements) which is non-native to many experimental platforms. Nonetheless, it was demonstrated in a number of experiments \cite{Struchalin2021,shadow_ion_experiment,ZhangPRL21}.


Besides the information theoretic complexity proofs \cite{Aaronson2007,aaronson2018shadow,Huang2020,Rocchetto2017} the quest for designing a practical protocol without computational bottlenecks is still ongoing. Here the tensor network representation of quantum states comes in handy. The tensor network approach is extremely successful in modeling quantum systems of large dimensions \cite{Waintal, ayral2023density, Continious_Dynamics_Intro_Orus2019, Open_Quantum_Dynamics_Intro_PhysRevLett.122.160401, quinones2019tensor, orus2019tensor,kurmapu2022reconstructing}. It had recently found application for full QST \cite{Cramer2010, Lanyon2017}. 
It is based on a constructive proof \cite{PhysRevLett.111.020401} that generic matrix product operators are completely determined by their local reductions. The low-rank tensor representation is actively used for tomography of quantum processes \cite{PhysRevLett.130.150402, yu2023learning, dang2021process, Torlai2023}. The work \cite{Akhtar2023scalableflexible} demonstrates an interesting method combining shadow tomography with local scrambling and tensor network methods.

Here we demonstrate a reconstruction algorithm based on the shadow tomography paradigm which requires only factorized single-qubit  measurements and is numerically efficient. The algorithm efficiently learns the specific form of a tensor network representation of a quantum state. We investigate the conditions when the algorithm follows the linear $O(N)$ dataset scaling and provide the in-depth analysis of the algorithm accuracy and convergence.

\section{\label{sec:OurApproach}The tensor network model}

\subsection{\label{sec:TTApproach}Tensor-Train representation of quantum states}

\begin{figure}[!t]
\includegraphics[width=1.0\linewidth]{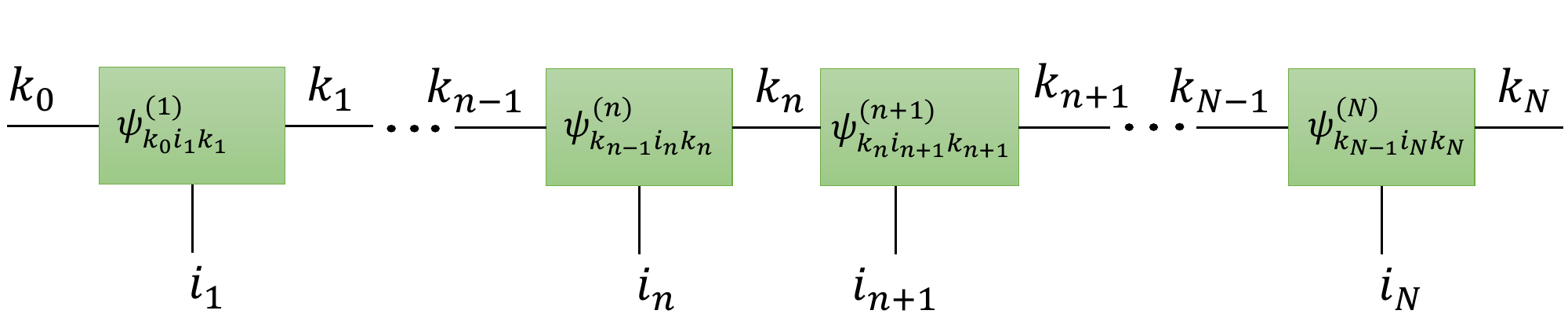}
\caption{\label{fig:tensor_train} Tensor train decomposition (\ref{Tensor_Train_state}) of tensor $\psi_{i_1 \dots i_N}$ describing the amplitudes expansion of the quantum state (\ref{Superposition_N_qubits_state}) in terms of the basic states of the computational basis of the $N$-qubit system. The green squares denote the TT-cores of the tensor train, the lines connecting the green squares denote the indices over which summation is performed, and the lines with free ends that point downwards denote the physical indices $i_n$ of the original tensor $\psi_{ i_1 \dots i_N }$.}
\end{figure}

An arbitrary quantum state $|\psi\rangle$ of $N$ qubits can be expanded into the basic states of the system:

\begin{equation}\label{Superposition_N_qubits_state}
    |\psi\rangle = \sum_{i_1, \dots, i_N = 0}^{1} \psi_{i_1 \dots i_N} |i_1\rangle \otimes \dots \otimes |i_N\rangle,
\end{equation}
where $\psi_{i_1 \dots i_N}$ is a tensor with complex coefficients. The full $\psi_{i_1 \dots i_N}$ tensor description quickly becomes intractable with increasing $N$. The {\it tensor-train (TT) decomposition} \cite{doi:10.1137/090752286} of the tensor $\psi_{i_1 \dots i_N}$ is an expansion (see Fig.~\ref{fig:tensor_train} for graphical representation)

\begin{equation}\label{Tensor_Train_state}
    \psi_{i_1 \dots i_N} = \sum_{k_0, \dots, k_N = 1}^{r_0, \dots, r_N} \psi_{k_0 i_1 k_1}^{(1)} \dots \psi_{k_{N - 1} i_N k_N}^{(N)},
\end{equation}
where the tensors $\psi_{k_{n - 1} i_n k_n}^{(n)}$ are called {\it TT-cores} and the values $r_n$ are called {\it TT-ranks} of the tensor train.  Such a decomposition is also known in physics as the {\it matrix product state} (MPS) \cite{schollwock2011density, perez2006matrix, zaletel2015time}. The TT decomposition provides a computationally efficient description of a quantum state in case of low-complexity states, for instance, generated by a low-depth quantum circuit \cite{Waintal, ayral2023density}.

The efficiency of this decomposition in terms of the amount of memory required to store TT-cores $\psi_{k_{n - 1} i_n k_n}^{(n)}$ depends on the value of TT-ranks $r_n$, which are determined by the ranks of {\it unfolding matrices} $\psi_n$ of $\psi_{i_1 \dots i_N}$, which are obtained from the tensor $\psi_{i_1 \dots i_N}$ by reshaping using index grouping -- the first $n$ indices enumerate the rows of $\psi_n$, and the last $N - n$ the columns of $\psi_n$ \cite{doi:10.1137/090752286}.

In a general case a mixed state $\hat{\rho}$ of an $N$-qubit register is expanded as
\begin{equation}\label{Superposition_N_qubits_operator}
    \hat{\rho} = \sum_{\substack{i_1, \dots, i_N = 0,\\ j_1, \dots, j_N = 0}}^{1} \rho_{i_1 \dots i_N j_1 \dots j_N} |i_1\rangle\langle j_1| \otimes \dots \otimes |i_N\rangle\langle j_N|,
\end{equation}
where $\rho_{i_1 \dots i_N j_1 \dots j_N}$ is $2N$-dimensional tensor. For $\rho_{i_1 \dots i_N j_1 \dots j_N}$, we can write the expansion in terms of the basis in the following form:
\begin{equation}\label{Tensor_Train_operator}
    \rho_{i_1 \dots i_N j_1 \dots j_N} = \sum_{k_0, \dots, k_N = 0}^{r_0, \dots, r_N} \rho_{k_0 i_1 j_1 k_1}^{(1)} \dots \rho_{k_{N - 1} i_N j_N k_N}^{(N)}.
\end{equation}
This representation is also known as a matrix product operator (MPO) form \cite{chan2016matrix}.

\subsection{\label{sec:StateMeasModel}State and measurement model}

We design a TT-based protocol which implements a learning principle described in \cite{Aaronson2007, AaronsonOnlineLearning}. The end-goal of the protocol is a model $\sigma$ of an unknown quantum state $\hat{\rho}$. The learning starts with a training dataset $\{\hat{E}_{m}^{t},p_{m}^{t}\}$ where $\hat{E}_{m}^{t}$ is a projective single-qubit measurement and the $p_{m}^{t}$ is the probability of acceptance for this measurement. The protocol then uses the training dataset to construct a plausible model $\sigma$.  We start by describing a mathematical model of the state $\hat{\rho}$ and the measurement $\hat{E}$ which we use in our simulation.

\begin{figure}[!t]
\includegraphics[width=0.5\linewidth]{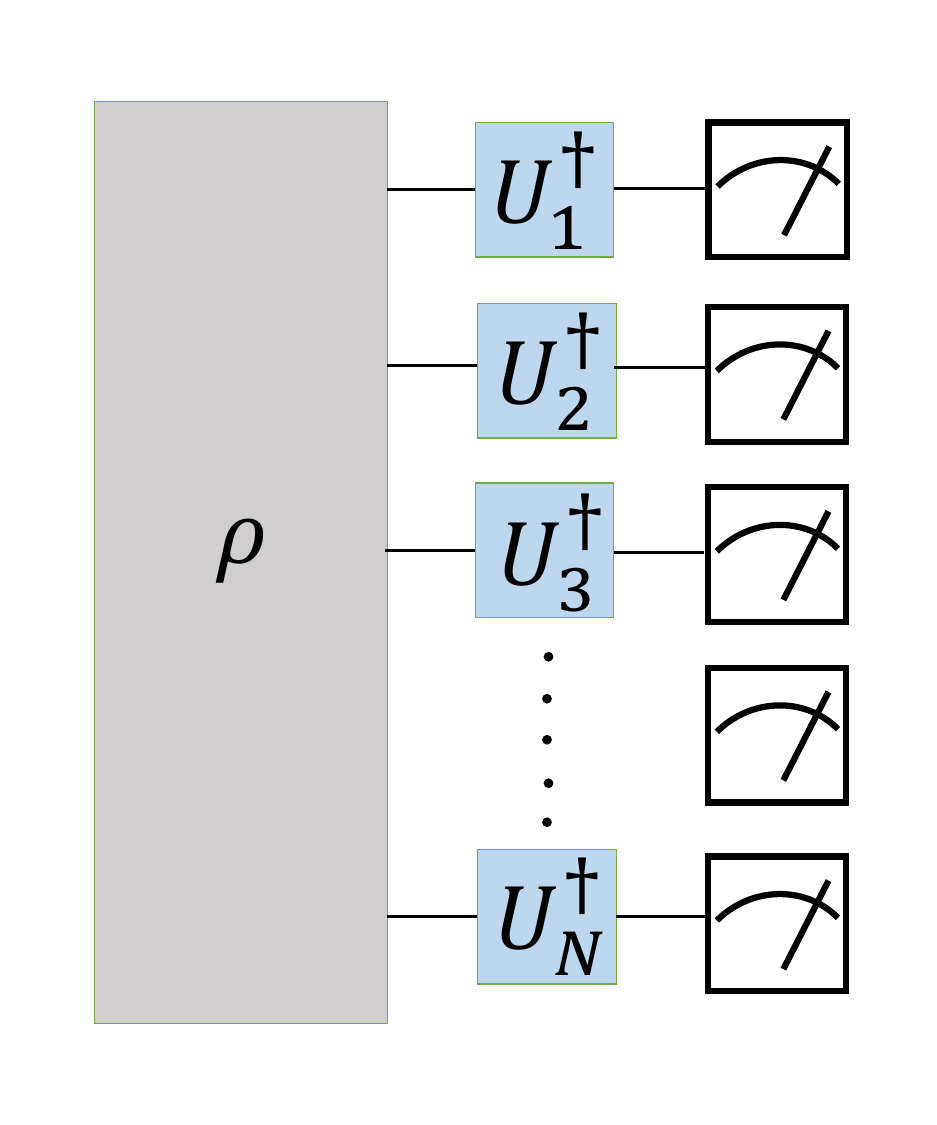}
\caption{\label{fig:scheme_meas} The scheme for generating the training dataset. The blue squares represent random single-qubit gates contained in the formula (\ref{E_operator_meas_with_U}).}
\end{figure}

The generic mixed state $\hat{\rho}$ is represented as an MPO tensor network (\ref{Tensor_Train_operator}). The measurement applied to the register is mathematically described by a positive operator valued measure (POVM) $\{\hat{E}^{(j)}\}$, where each $\hat{E}^{(j)}$ is a Hermitian-positive semidefinite operator such that $\sum_{j} \hat{E}^{(j)} = \hat{I}$. The probability of a measurement outcome $j$ is $p_j = \operatorname{Tr}(\hat{E}^{(j)}\hat{\rho})$. We will focus on two-outcome POVMs $\{\hat{E}^{(1)} = \hat{E},  \hat{E}^{(2)} = \hat{I} - \hat{E}\}$ due to the existence pf the learnability proof \cite{aaronson2018shadow} and their direct implementation in the experiment \cite{rocchetto2019experimental}. We consider only projective single-qubit measurements, that is, we use only measurements described by tensor products of single-qubit measurement operators:
\begin{equation}\label{Measurement_operators}
    \hat{E} = \hat{E}_1 \otimes \dots \otimes \hat{E}_N,
\end{equation}
moreover, here $\hat{E}_i = |\psi_i\rangle\langle\psi_i|$ is a projector onto a pure state $|\psi_i\rangle$. That is, in a tensor train representation, such an operator can be represented by the tensor train with the maximum rank $\chi = 1$. 
We choose the measurement operators $\hat{E}$ in the form:
\begin{equation}\label{E_operator_meas_with_U}
    \hat{E} = \hat{U}_1|0\rangle\langle 0|\hat{U}_1^{\dagger} \otimes \dots \otimes \hat{U}_N|0\rangle\langle 0|\hat{U}_N^{\dagger},
\end{equation}
where the unitary operators $\hat{U}_k = \hat{R}_{\vec{n}}(\theta)$ are the rotations of the Bloch sphere around a random unit vector $\vec{n}$ by the angle $\theta$. The unit vector $\vec{n}$ is uniformly distributed over the $4\pi$ solid angle, and the angle $\theta$ is uniformly distributed from $0$ to $2 \pi$. Fig.~\ref{fig:scheme_meas} illustrates the simulated setup which outputs the training dataset.

The random mixed state sampling procedure in the MPO format begins with generating a random operator $\hat{L}$ immediately in the MPO format with all ranks $r_i$ having the same $\chi_S$ value, except for the boundary ones, which have a rank of $1$ ($r_0 = r_N = 1$). Each TT-core element is complex and has the form $l^{(n)}_{k_{n - 1} i_n j_n k_n} = x^{(n)}_{k_{n - 1} i_n j_n k_n} + i y^{(n)}_{k_{n - 1} i_n j_n k_n}$, where $x^{(n)}_{k_{n - 1} i_n j_n k_n}, y^{(n)}_{k_{n - 1} i_n j_n k_n}$ are random variables distributed according to the standard normal probability distribution $\mathcal{N}(0, 1)$. We then use a contraction algorithm to calculate $\operatorname{Tr}\hat{L}\hat{L}^{\dagger}$ and output the MPO density operator:
 \begin{equation}\label{DensMatrixTest}
    \hat{\rho} = \dfrac{\hat{L}\hat{L}^{\dagger}}{\operatorname{Tr} (\hat{L}\hat{L}^{\dagger})}.
\end{equation}

It is easy to see that this choice of $\hat{\rho}$ satisfies all the requirements for the density operator: $\hat{\rho}^{\dagger} = \hat{\rho}$, $\hat{\rho} > 0$, $\operatorname{Tr} \hat{\rho} = 1$. The resulting random density operator is represented in TT format with unit boundary ranks, and all other ranks equal to $\chi_S^2$.

Next, we sample random density operator $\hat{\rho}$ by the formula (\ref{DensMatrixTest}), generate $M$ different measurement operators $\hat{E}_m$ by the formula (\ref{E_operator_meas_with_U}), and calculate the probabilities $p_m = \operatorname{Tr} (\hat{E}_m \hat{\rho})$. As a result, we get a training set of size $M$ which is then fed to the learning protocol as a tuple $\{\hat{E}_m, p_m\}$.

\subsection{\label{sec:LearningModel}Learning model}

The learning process seeks for the model $\sigma$ which not only replicates the training data $\{\hat{E}_m^{t}, p_m^{t}\}$ but has the power to predict the success probability $\tilde{p}$ of a new measurement $\hat{E} \notin \{\hat{E}_m^{t}, p_m^{t}\}$. We express the model $\sigma$ as
\begin{equation}\label{Learning_model}
    \tilde{p} = \operatorname{Tr} (\hat{E} \hat{\sigma}) =\operatorname{Tr} (\hat{E} \hat{\Omega} \hat{\Omega}^{\dagger}).
\end{equation}

\begin{figure}[!t]
\includegraphics[width=1.0\linewidth]{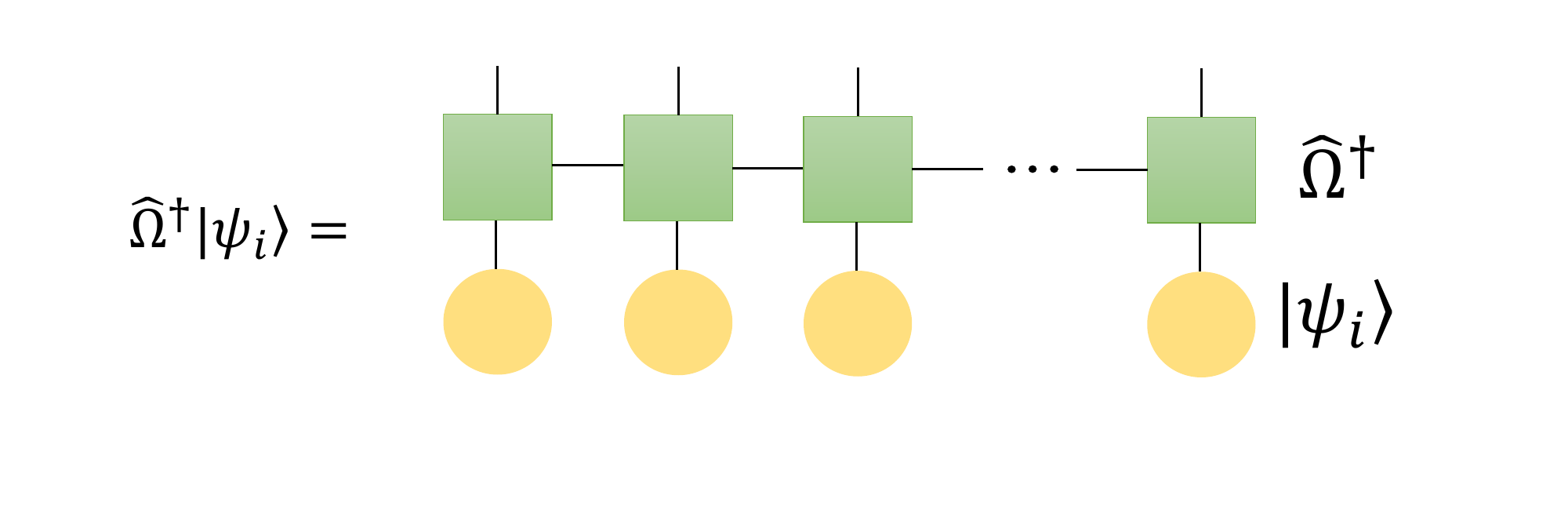}
\caption{\label{fig:LearningProcedure} Graphic representation of the procedure for calculating the probability by the formula (\ref{TraceCalculation}) of obtaining a positive result in an experiment with the POVM operator $\hat{E_i} = |\psi_i\rangle\langle\psi_i|$ in the tensor train format. The yellow circles represent the MPS $|\psi_i\rangle$ TT-cores, and the green squares represent the MPO $\hat{\Omega}^{\dagger}$ TT-cores. The lines that connect the TT-cores indicate that summation is performed on the corresponding indices. The yellow circles have only one line because $\hat{E_i}$ has all TT-ranks equal to 1.}
\end{figure}

Since $\hat{E}$ is a projector onto a pure state $|\psi\rangle$, the expression (\ref{Learning_model}) can be immediately  rewritten in a form that is more convenient for calculations:
\begin{eqnarray}\label{TraceCalculation}
    \tilde{p} = &&\operatorname{Tr} (\hat{E} \hat{\Omega} \hat{\Omega}^{\dagger}) = \operatorname{Tr} (\hat{\Omega}^{\dagger} \hat{E} \hat{\Omega}) = \operatorname{Tr} (\hat{\Omega}^{\dagger} |\psi\rangle\langle\psi| \hat{\Omega}) = \nonumber\\
    &&= \langle\psi|\hat{\Omega}\hat{\Omega}^{\dagger}|\psi\rangle = ||\hat{\Omega}^{\dagger}|\psi\rangle||^2.
\end{eqnarray}
The algorithm to calculate the probability $\tilde{p}$ in tensor network terms is illustrated in Fig.~\ref{fig:LearningProcedure}.

The operator $\hat{\Omega}$ allows the expansion (\ref{Superposition_N_qubits_operator}) with  $\Omega_{i_1 \dots i_N j_1 \dots j_N}$ being the coefficient tensor. We restrict $\Omega_{i_1 \dots i_N j_1 \dots j_N}$ to be a tensor train with a maximal individual TT-rank $\max r_i = \chi_M$ . Moreover, we assume that the boundary TT-ranks are equal to one, and the ranks of the rest of the cores are equal to the maximal rank $\chi_M$ ($r_0 = r_N = 1$ and $r_k = \chi_M$ for $k \not= 1, N$):

\begin{equation}\label{Tensor_Train_omega}
    \Omega_{i_1 \dots i_N j_1 \dots j_N} = \sum_{k_1, \dots, k_N = 0}^{r_0, \dots, r_N} \omega_{k_0 i_1 j_1 k_1}^{(1)} \dots \omega_{k_{N - 1} i_N j_N k_N}^{(N)}.
\end{equation}

The representation of the density operator $\hat{\Omega} \hat{\Omega}^{\dagger}$, where $\hat{\Omega}$ is a tensor train is called the {\it locally purified state} (LPS) \cite{Torlai2023}.

The real and imaginary parts of TT-cores $\omega_{k_{n - 1} i_n j_n k_n}^{(n)}$ are the learning parameters in our setting. We optimize the parameters of the $\hat{\Omega}$  (\ref{Learning_model}) to reach the most accurate prediction of the success probabilities $\tilde{p}$ of a factorized measurement. We solve an archetypal supervised learning problem: the model parameters are optimized based on the training dataset $\{\hat{E}_m^{t}, p_m^{t}\}$ of size $M$ and the quality of the model is evaluated on an independent  test set $\{\hat{E}_n, p_n\} \notin \{\hat{E}_m^{t}, p_m^{t}\}$.

\begin{figure*}[!t]
\begin{minipage}{0.3\linewidth}
\center{\includegraphics[width=1.09\linewidth]{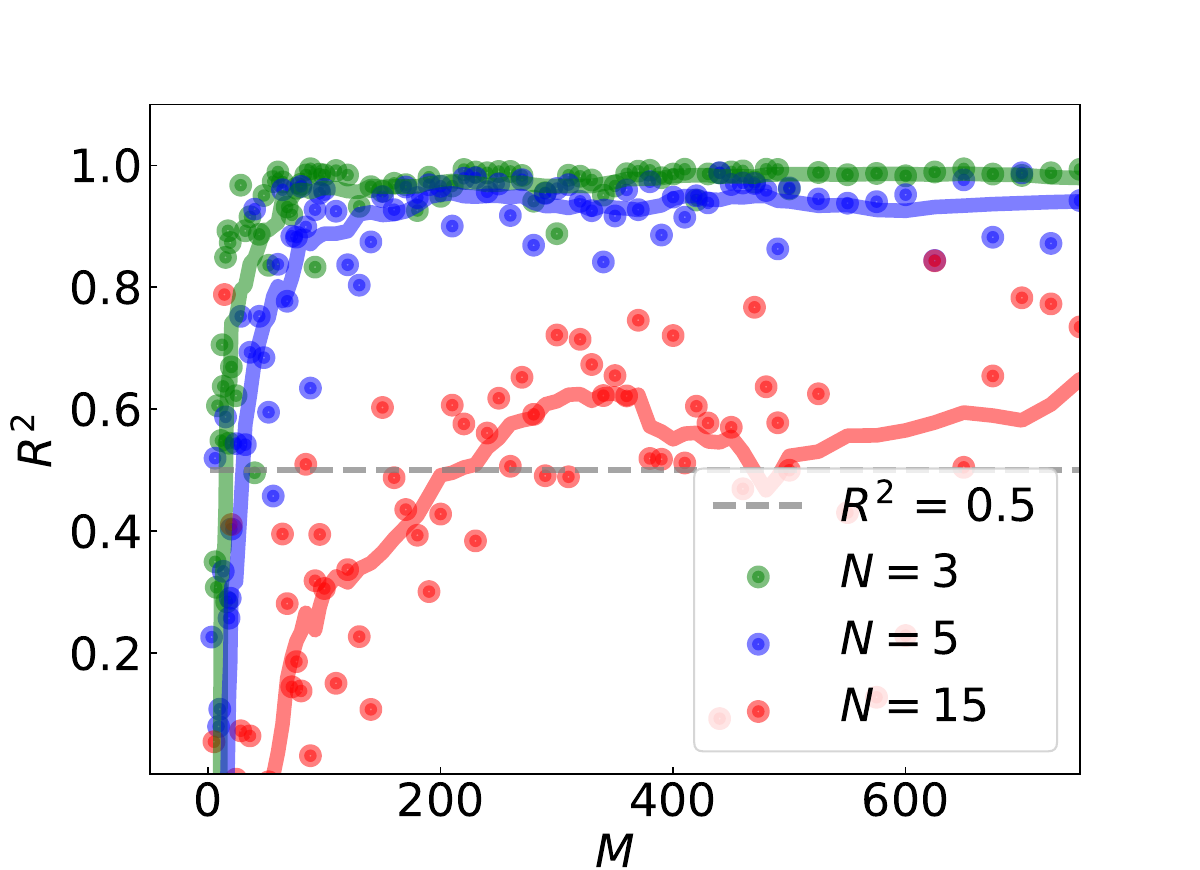}} a) $\chi_S = 2$, $\chi_M = 4$
\end{minipage}
\begin{minipage}{0.3\linewidth}
\center{\includegraphics[width=1.09\linewidth]{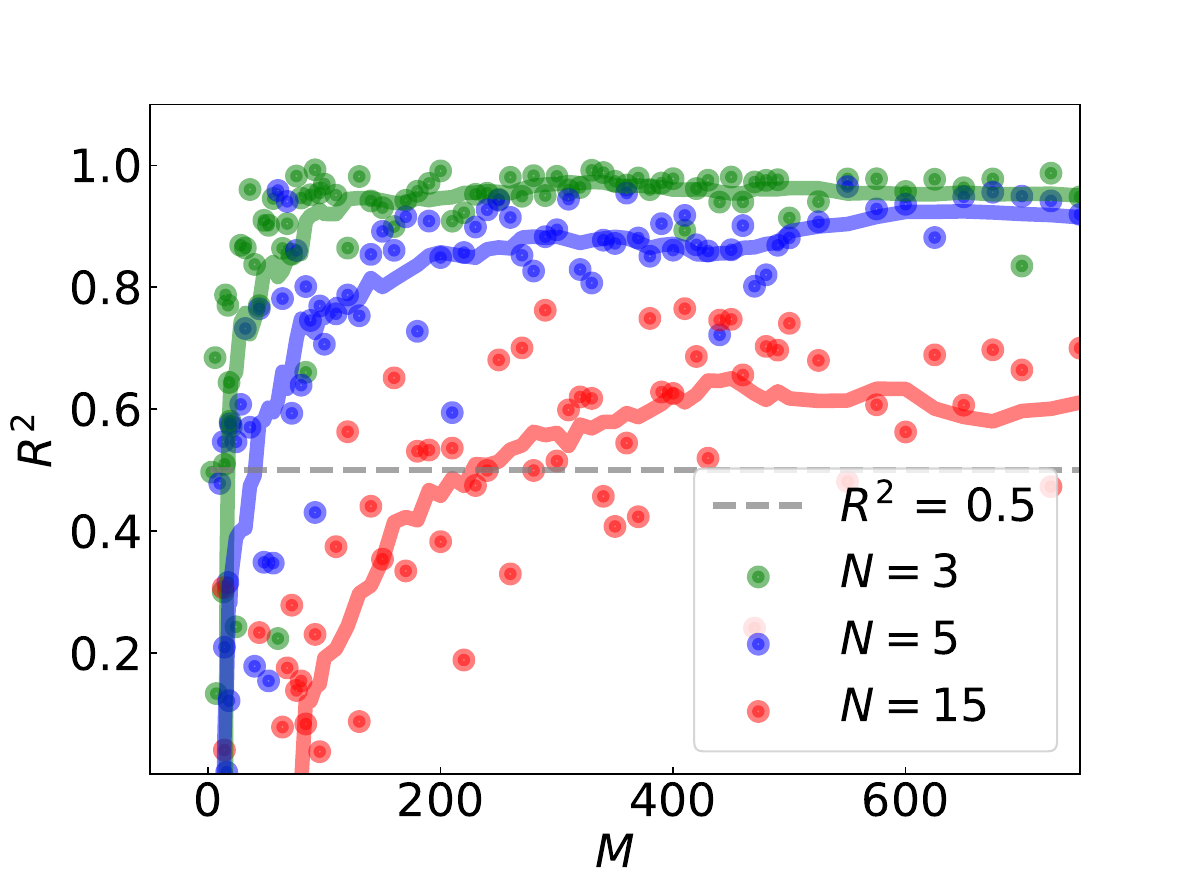}} b) $\chi_S = 4$, $\chi_M = 4$
\end{minipage}
\begin{minipage}{0.3\linewidth}
\center{\includegraphics[width=1.09\linewidth]{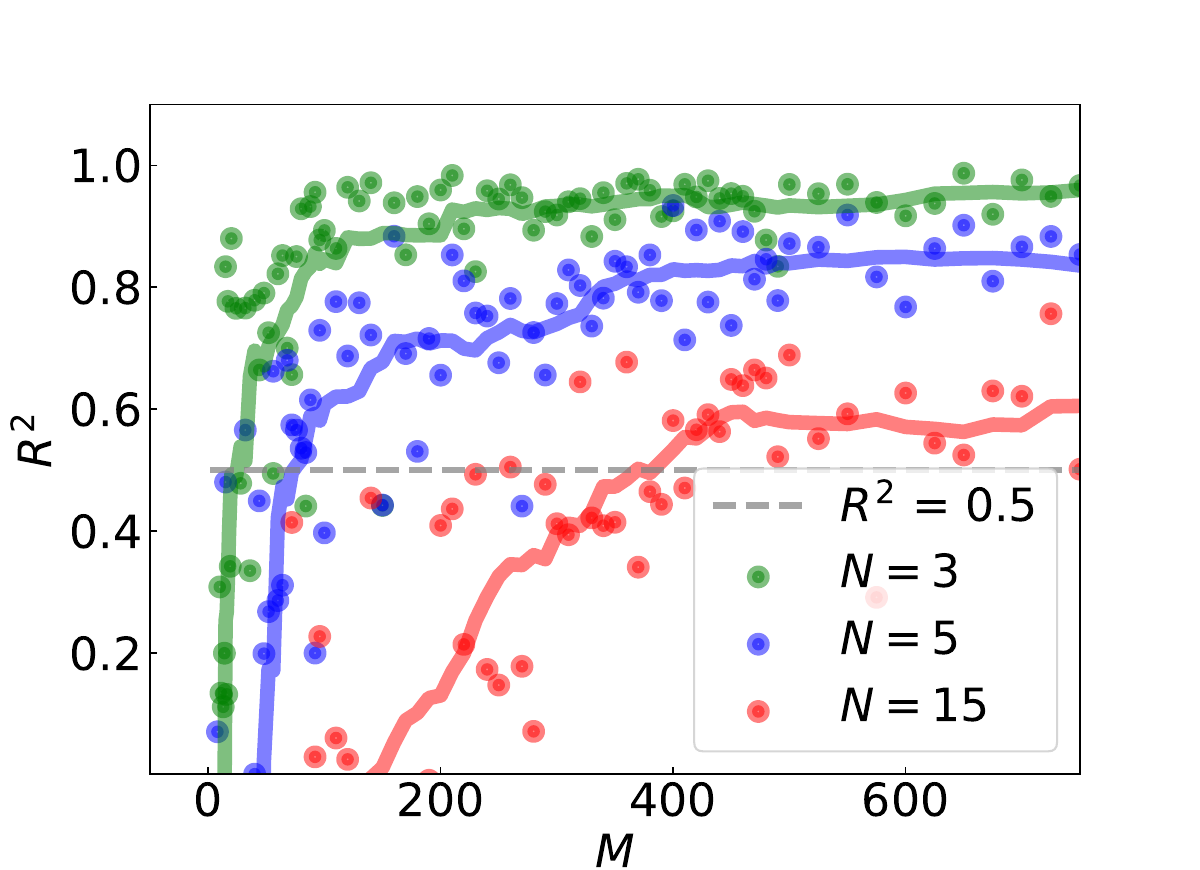}} c) $\chi_S = 8$, $\chi_M = 4$
\end{minipage}
\begin{minipage}{0.3\linewidth}
\center{\includegraphics[width=1.09\linewidth]{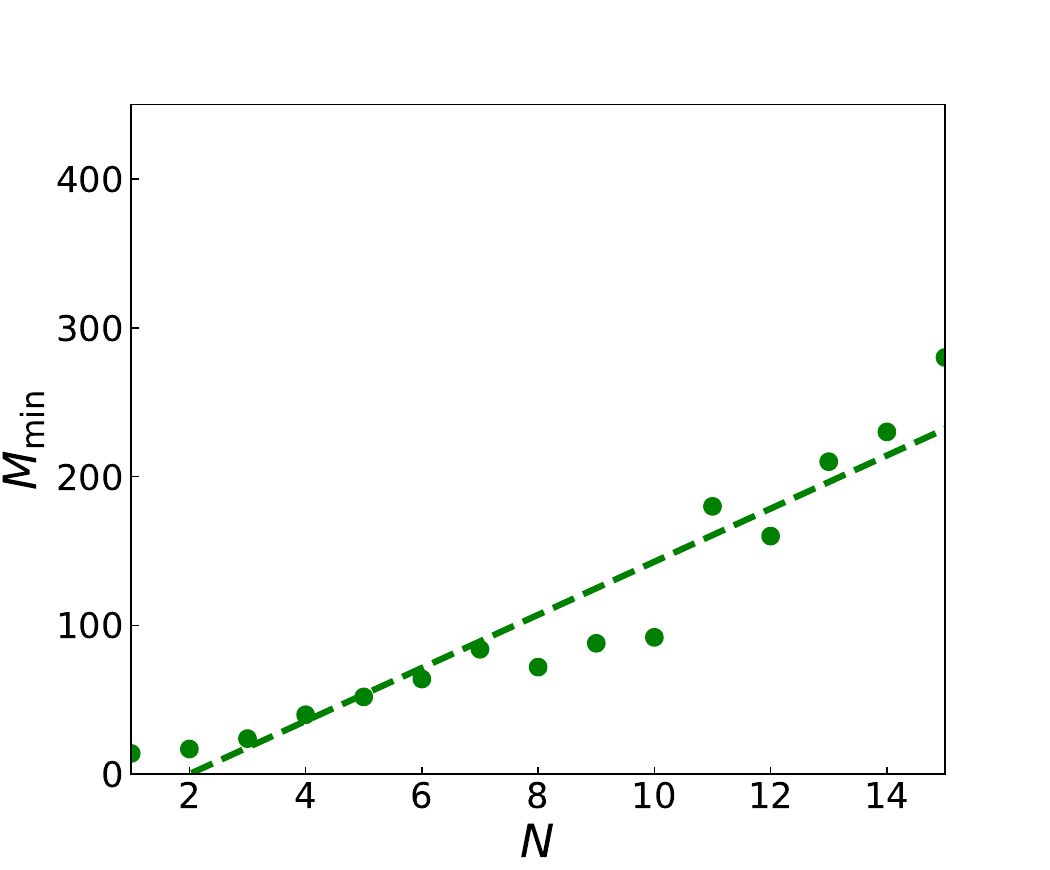}} d) $\chi_S = 2$, $\chi_M = 4$
\end{minipage}
\begin{minipage}{0.3\linewidth}
\center{\includegraphics[width=1.09\linewidth]{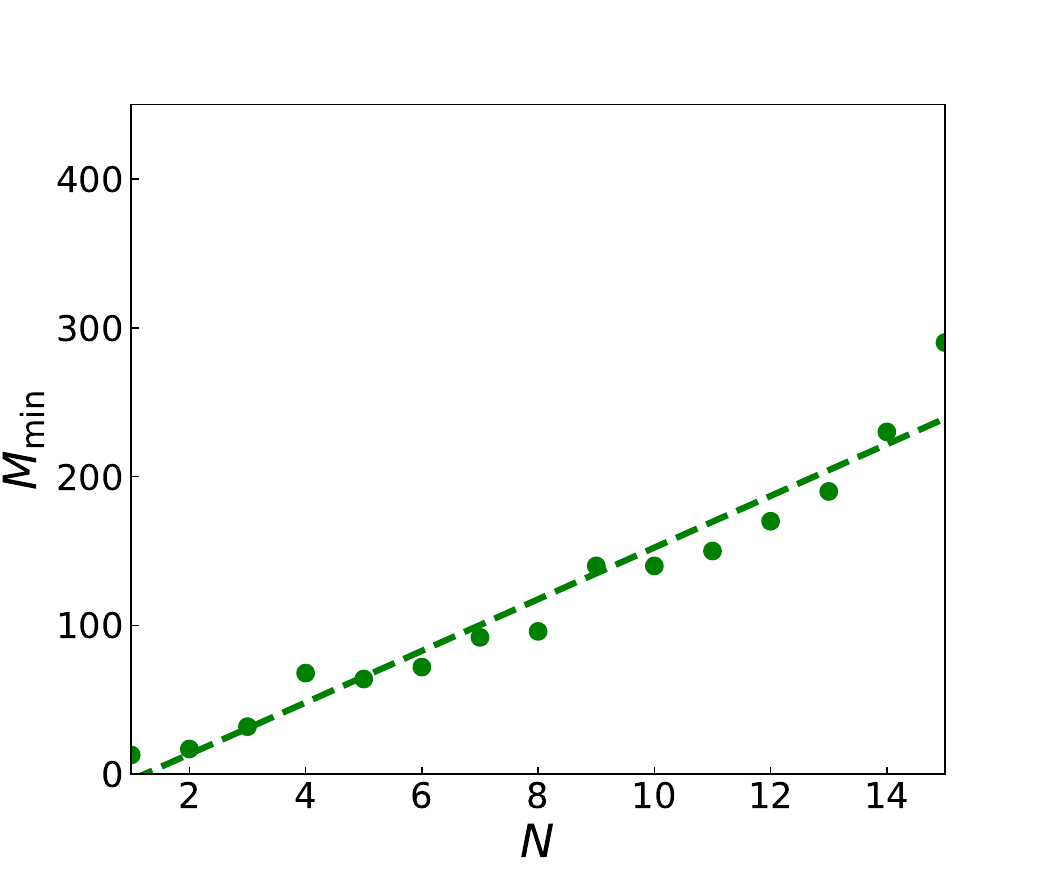}} e) $\chi_S = 4$, $\chi_M = 4$
\end{minipage}
\begin{minipage}{0.3\linewidth}
\center{\includegraphics[width=1.09\linewidth]{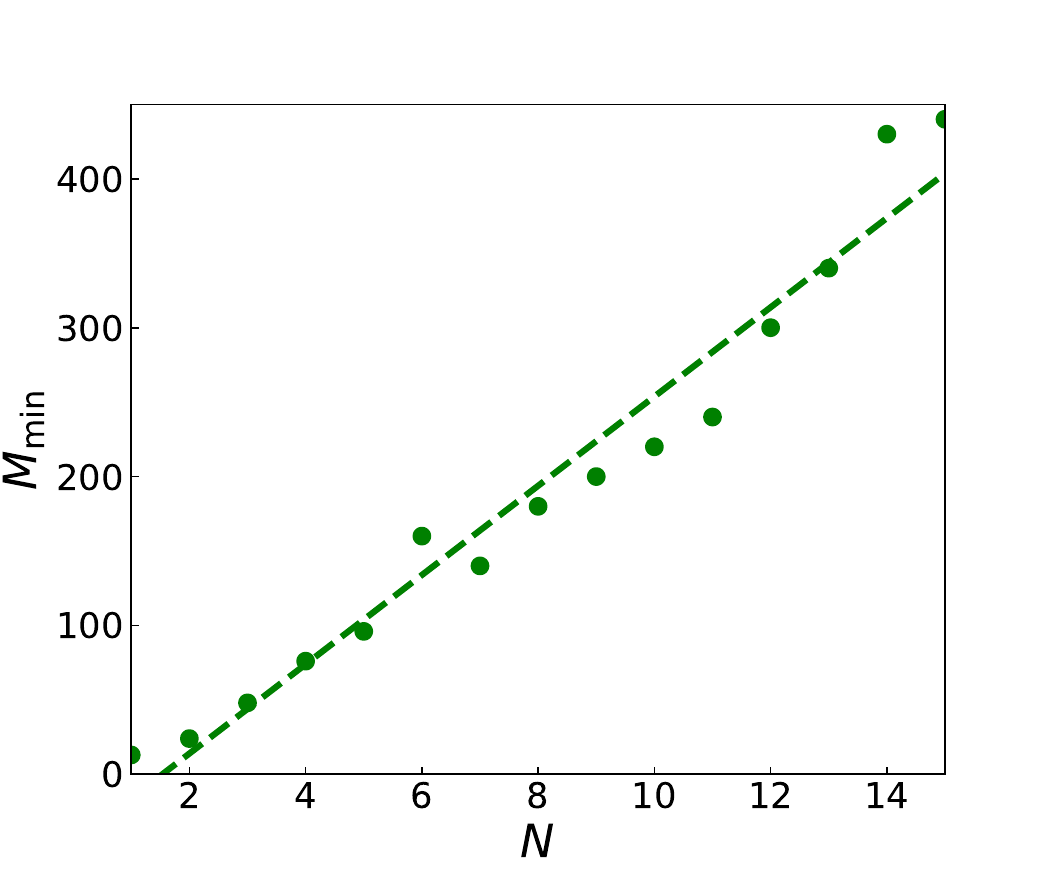}} f) $\chi_S = 8$, $\chi_M = 4$
\end{minipage}
\caption{\label{fig:NumResLinearM}Generalized results of numerical simulation of the tensor train training procedure. The upper plots (a, b, c) show how the key indicator of learning quality $R^2$ depends on the size of the training sample (points) and moving averages are plotted with a symmetrical window of 11 points (solid line). The lower plots (d, e, f) show the minimum value of the training sample size to achieve success in learning by the criterion $R^2 = 0,5$ on the test sample (for the moving average).}
\end{figure*}

Let us discuss the physical validity of choosing the model $\sigma=\hat{\Omega} \hat{\Omega}^{\dagger}$. This choice satisfies the condition that the probability $\tilde{p}=\operatorname{Tr}(\hat{E} \hat{\Omega} \hat{\Omega}^{\dagger})$ must be a real non-negative number. However, this does not guarantee completeness, namely, if we take a complete set of measurement operators that sum to one, then the sum of the corresponding probabilities will not necessarily be equal to one, that is, the trace of our simulated density matrix is not necessarily equal to one. We note that the original paper \cite{Aaronson2007} speculated that $\sigma$ does not necessarily have to represent a physically meaningful object. This fact stems from the assumption that we no longer look for the description of the unknown state $\hat{\rho}$ and are only interested in predicting particular features associated with this state.

\subsection{\label{sec:LossFunc}Loss function and performance criterion}

We optimize the parameters of $\hat{\Omega}$ by minimizing the mean squared error (MSE) loss function:
\begin{equation}\label{LossFuncFormula}
    \mathcal{L} = \dfrac{1}{M} \sum_{m = 1}^{M} (\operatorname{Tr}(\hat{E}_m \hat{\Omega} \hat{\Omega}^{\dagger}) - p_m)^2.
\end{equation}
We use the normalization factor $C = (2^N)^2$, that is $\mathcal{L}' = C \mathcal{L}$ to compensate for the $\sim \dfrac{1}{2^N}$ declining average values of the probabilities $\tilde{p}$ in the larger dimensions. We supplement the loss function value $\mathcal{L}'$ with the coefficient of determination $R^2$ which provides a comprehensible quality estimator:
\begin{equation}\label{R2}
    R^2 = 1 - \dfrac{\mathcal{L}}{\sigma^2_{p}},
\end{equation}
where $\sigma^2_{p}$ is the probability variance on the test sample. The value of $R^{2}$ shows how much the trained model outperforms a trivial model that returns a constant value of the probability $\bar{p}_n$ equal to the average value of the probabilities for all measurement operators $\hat{E}_n $ from the test set.

\section{\label{sec:NumericalResults}Numerical results}

The numerical experiment begins with sampling of the quantum state subject to the learning protocol using the formula (\ref{DensMatrixTest}) with the maximum TT-rank $\chi_S^2$. Then we initialize the model $\sigma$ using the same method (\ref{DensMatrixTest}) with the maximum TT-rank $\chi_M^2$. We consider cases where $\chi_M < \chi_S$, $\chi_M = \chi_S$ and $\chi_M < \chi_S$. We use the stochastic gradient descent algorithm to minimize the loss function (\ref{LossFuncFormula}). At each step the loss function value is averaged over the subset of $m = 5$ elements randomly drawn from the training dataset. This technique produces equally valid results as the averaging over the whole training dataset at each step and substantially reduces the computation cost.

Fig.~\ref{fig:NumResLinearM} shows the $R^2$ metric of the trained model after 1000 optimization steps. $R^{2}$ is calculated using an independent test dataset. Both training and test datasets have equal size $M$. We use the $R^2 = 0.5$ value as a threshold indicating that the model reached adequate performance. The exact value of this threshold is always ambiguous and does not affect the training dataset size scaling properties that are of the most interest. However, in Fig.~\ref{fig:NumResLinearM}(a, b, c)  we observe the $R^2$ saturation behaviour peaking at a certain level. The saturated $R^2$ value drops down in case of the larger number of qubits, which we associate with the imperfection of the non-convex multivariate optimization of the loss function with complex landscape. The $R^{2}$ saturation is also the consequence of the model $\sigma$ maximal rank being artificially limited and hence being in principle unable to render any property of the unknown state $\rho$ (see Appendix \ref{AppendixR2Constraint} and Appendix \ref{AppendixFullMapR2} for details). The PAC theory states that the model $\sigma$ will produce the outliers with low probability and we speculate that this might also be the reason of the $R^{2}$ plateau. For details, see Appendix \ref{AppendixProbabilityErrors}, in which we calculated for each case in Fig.~\ref{fig:ProbErrorDistr}(a, b, c) the sample standard deviation $\sigma_{\varepsilon_p}$ to estimate the proportion of strong outliers and calculated the proportion of cases where $|\varepsilon_p| > 3\sigma_{\varepsilon_p}$: a) $2.2\%$, b) $1.9\%$, c) $1.0\%$. The dependencies of the minimum size of the training dataset required to achieve the $R^2 \geq 0.5$ value are shown in Fig.~\ref{fig:NumResLinearM}(d, e, f). We performed tests in the three cases -- $\chi_{S}<\chi_{M}$, $\chi_{S}=\chi_{M}$ and $\chi_{S}>\chi_{M}$. The simulation results proved the evidence that the $\sigma$ model could be less descriptive than a true quantum state model and still yield accurate $\tilde{p}$ predictions. In each of three cases we observe the linear scaling of the training dataset size $M$ required for  the successful learning of $\sigma$ for the $N$-qubit state.

\begin{figure}[!t]
\includegraphics[width=1.0\linewidth]{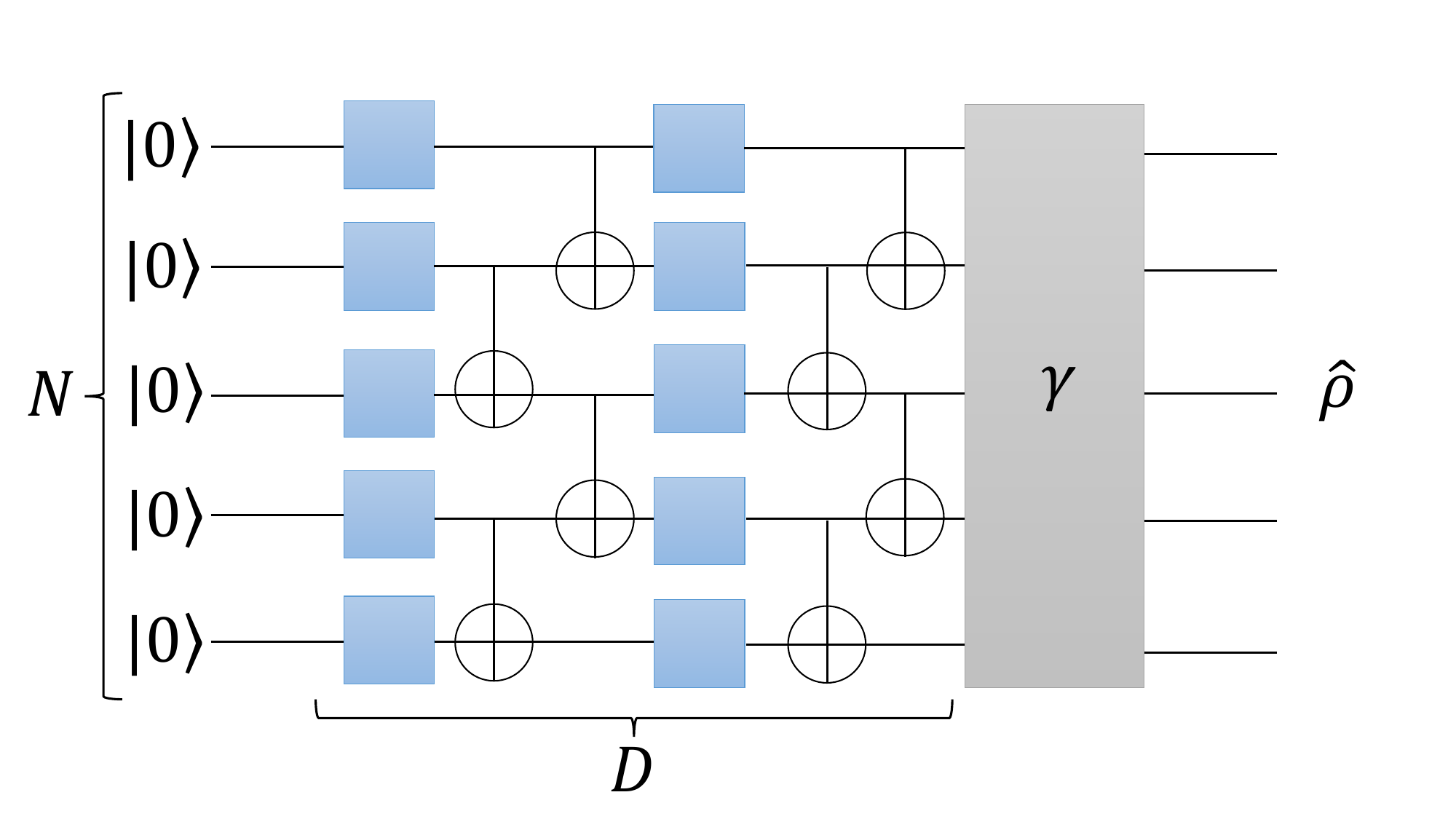}
\caption{\label{fig:phys_device_state} Standard hardware efficient ansatz used for NISQ hardware. The blue squares represent single-qubit rotations $\hat{U} = \hat{R}_{\vec{n}}(\theta)$ around a random unit vector $\vec{n}$ by the angle $\theta$. The unit vector $\vec{n}$ is uniformly distributed over the $4\pi$ solid angle, and the angle $\theta$ is uniformly distributed from $0$ to $2 \pi$. The gray rectangle at the end indicates depolarization channel (\ref{Depolarization_channel}) with the depolarization coefficient $\gamma$.}
\end{figure}

We validate our model using a realistic physical setting with well-known behaviour. Consider a quantum circuit shown in Fig.~\ref{fig:phys_device_state}, which generates a standard hardware efficient ansatz \cite{kandala2017hardware} ubiquitously used for NISQ hardware description. The low-depth circuits are well described by TT-based states with low-rank \cite{Waintal, ayral2023density} and should be efficiently learned with our protocol.

\begin{figure*}[!t]
\begin{minipage}{0.3\linewidth}
\center{\includegraphics[width=1.09\linewidth]{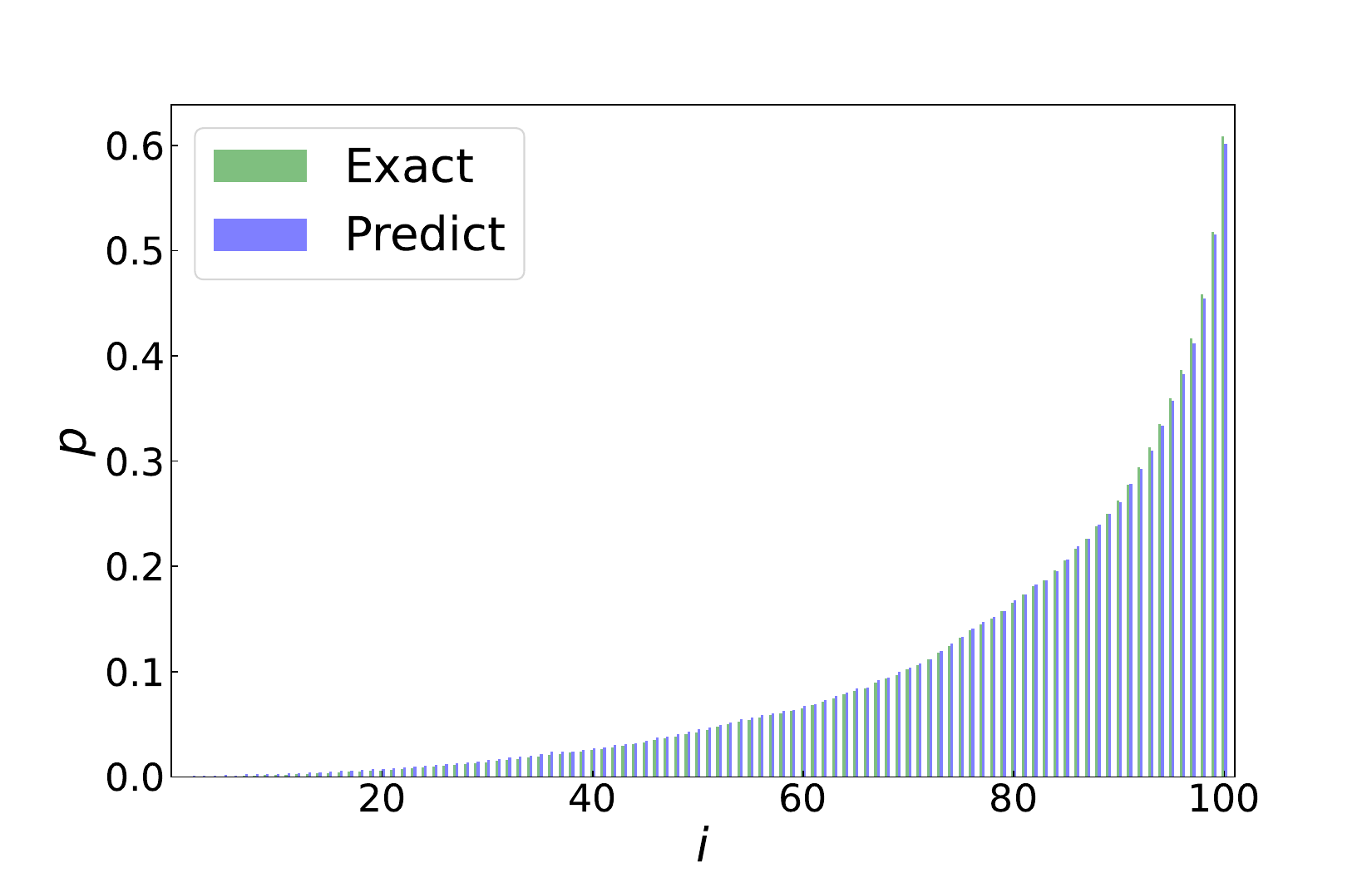}} a) $\gamma = 0,0$
\end{minipage}
\begin{minipage}{0.3\linewidth}
\center{\includegraphics[width=1.09\linewidth]{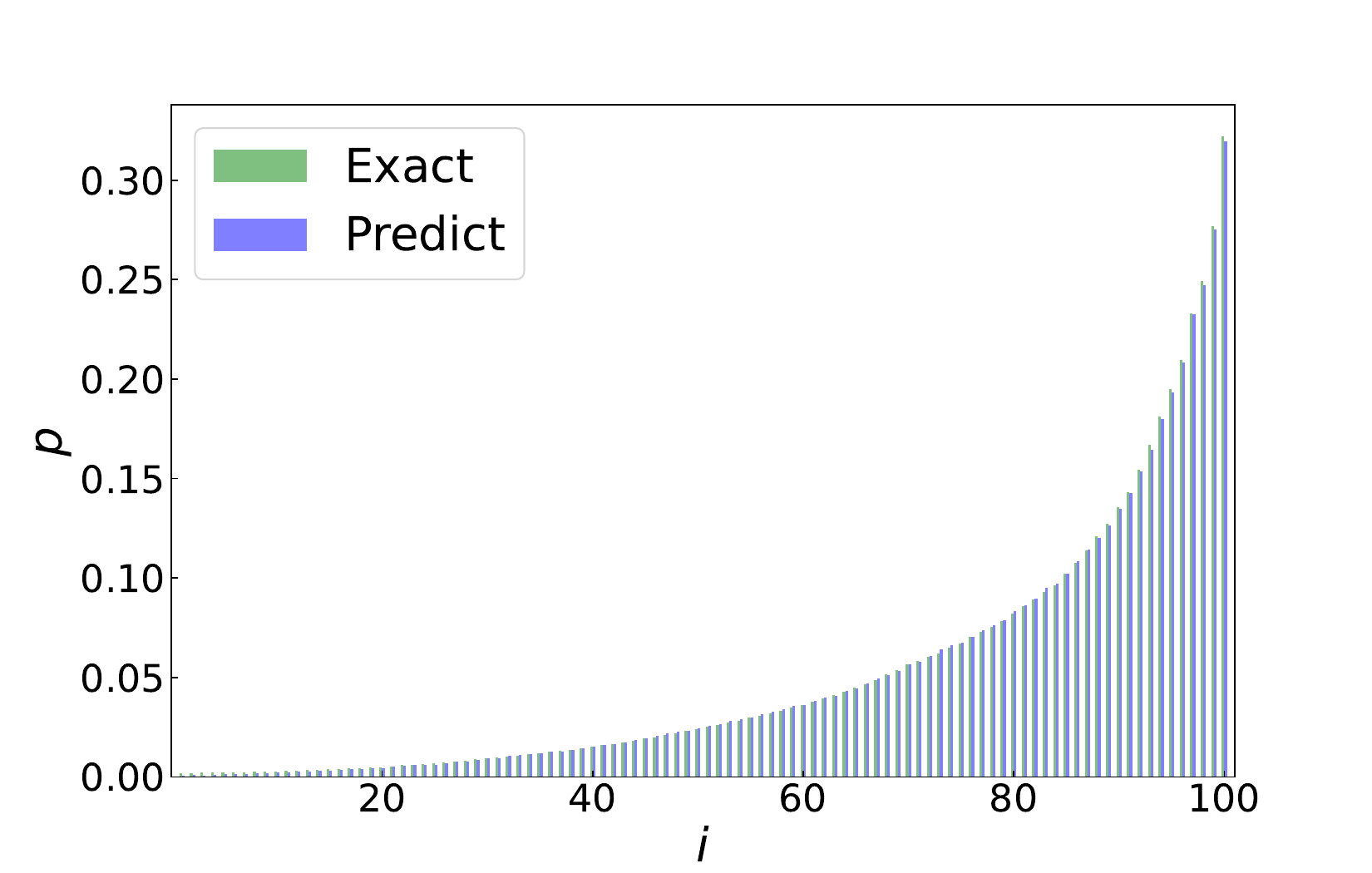}} b) $\gamma = 0,5$
\end{minipage}
\begin{minipage}{0.3\linewidth}
\center{\includegraphics[width=1.09\linewidth]{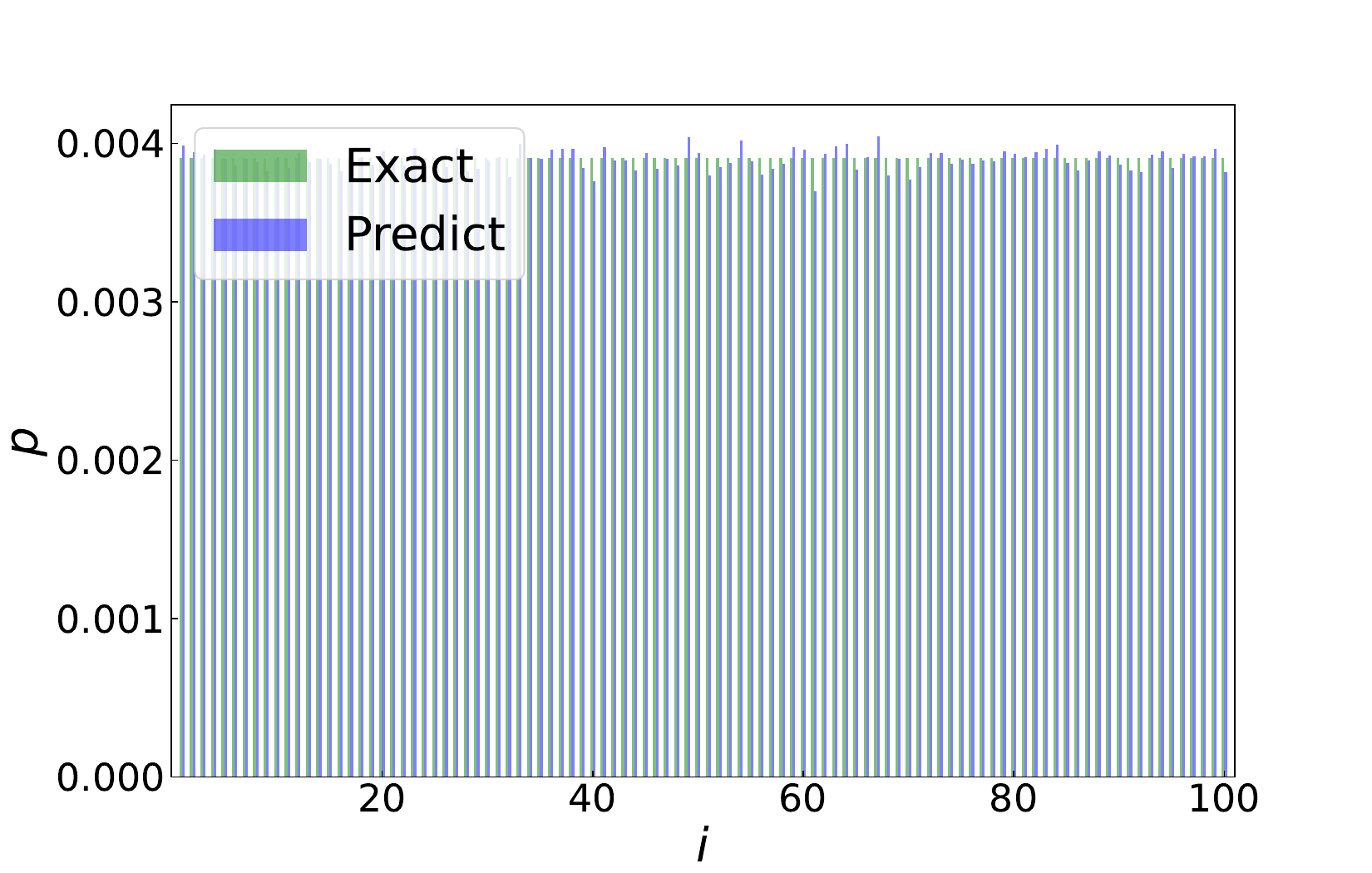}} c) $\gamma = 1,0$
\end{minipage}
\caption{\label{fig:NumResPhysDevice}Comparison of the exact probability distributions with the distributions generated by the trained model. The model was trained on the dataset of size $M=500$ and $5000$ optimization steps. The quantum state is the 8-qubit state $|0\rangle^{\otimes 8}$ at the output of the depth $D = 2$ random circuit supplemented with a depolarizing channel with a depolarization coefficient $\gamma$. Here $\chi_S^2 = 256$ and $\chi_M^2 = 16$. The results are shown with averaging over $30$ different sets of operators POVM $\hat{E}_i$.}
\end{figure*}

The initial quantum state $|0\rangle^{\otimes 8}$ passes through a random circuit of depth $D = 2$ consisting of two alternating layers of random single-qubit gates and entangling CNOT gates. In the general case, with a complete description of the quantum state at the output of such a circuit, the maximum TT-rank MPO will be equal to $\min(4^{[N / 2]}, 4^D) = 16$. Physical quantum systems are always subject to noise and their state, which should have been pure, becomes mixed. Therefore, afterwards the controlled level of noise is introduced by application of a depolarizing channel with a depolarization coefficient $\gamma$:
\begin{equation}\label{Depolarization_channel}
    \Delta_{\gamma} (\hat{\rho}) = (1 - \gamma) \hat{\rho} + \dfrac{\gamma}{2^N} \hat{I}.
\end{equation}
In our numerical example we used a circuit with $N=8$ qubits and the rank of the trained model was chosen to be $\chi_M^2 = 16$. 

The Fig.~\ref{fig:NumResPhysDevice} compares the probability distribution of the measurement outcomes predicted by the trained model $\sigma$ and the exact one. 

\begin{figure}[h]
\includegraphics[width=0.7\linewidth]{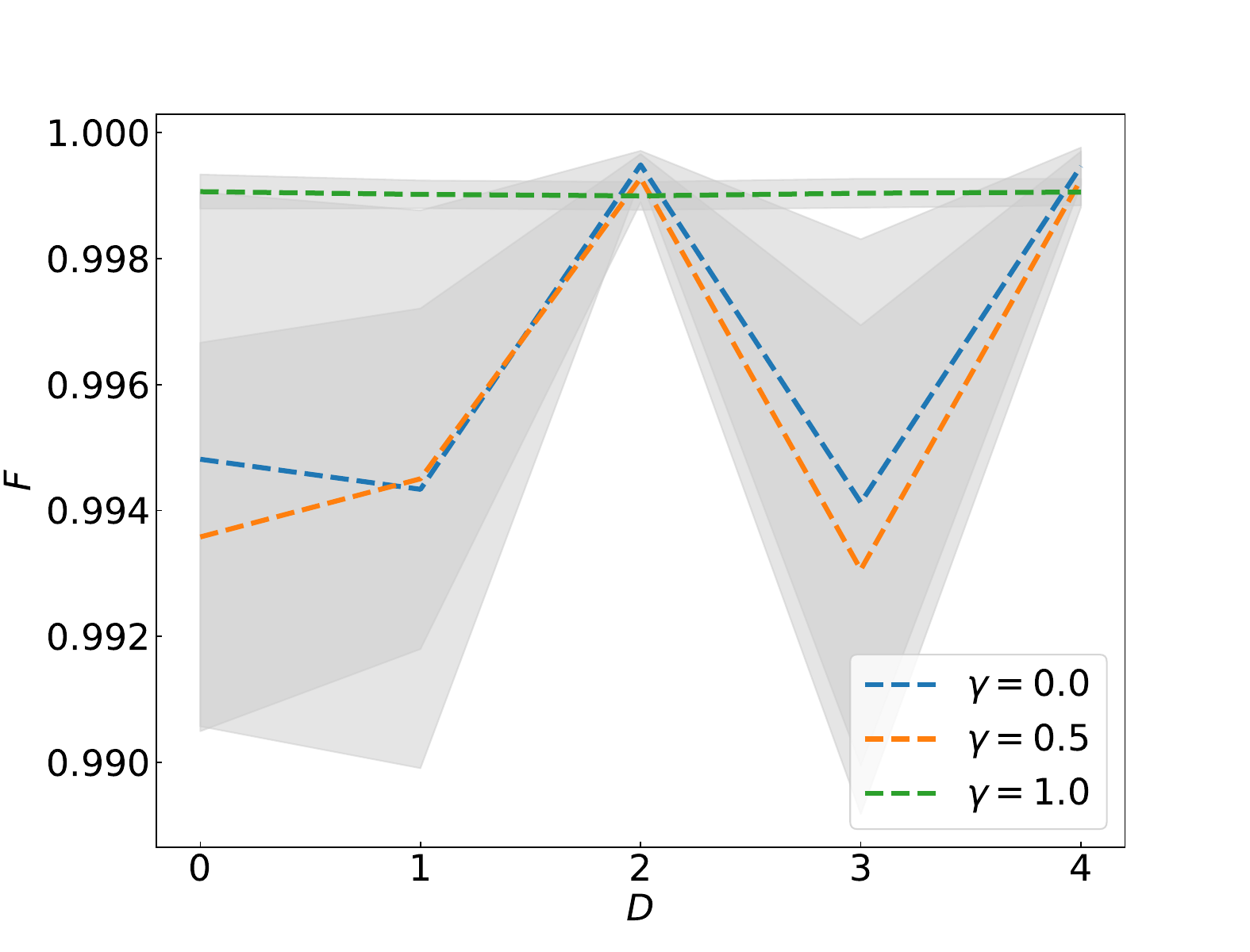}
\caption{\label{fig:fidelity_d} Fidelity $F$ between the probability distributions generated by the exact reference state and the trained model. The training dataset size ise $M = 500$, the number of optimization steps is $5000$ depending on the depth of the circuit $D$ and the depolarization coefficient $\gamma$.}
\end{figure}

Three different values of the depolarization coefficient $\gamma$ were used: $\gamma = 0$, corresponds to a pure state Fig.~\ref{fig:NumResPhysDevice}~a), $\gamma = 0.5$ producing a state with intermediate purity Fig.~\ref{fig:NumResPhysDevice}~b), and $\gamma = 1.0$ corresponds to a completely mixed state Fig.~\ref{fig:NumResPhysDevice}~c).  The indices $i$ of POVM-elements $\hat{E}_i$ are plotted along the horizontal axis and sorted by the absolute value of the exact probability. We observe qualitatively good correspondence between the model output and the exact values. The suitable quantitative figure of merit is fidelity between the two probability distributions:
\begin{equation}
    F = \dfrac{\sum_i \sqrt{p_i q_i}}{\sqrt{\sum_i p_i \sum_i q_i}}.
\end{equation}
We ran a series of tests for depths $D$ from $0$ to $4$ (full rank for $8$ qubits) and for three values of the depolarization coefficient, and found that the fidelity values ranged from $0.99$ to $0.999$ (Fig.~\ref{fig:fidelity_d}).

\section{\label{sec:Discussions}Conclusion}

We have demonstrated a practical algorithm for quantum state learning using a low-rank tensor-network model. Compared to classical shadows and similar approaches our algorithm uses only experimentally feasible single-qubit measurements and thus can be readily realized on the various existing experimental setups without any additional overhead. We provide numerical evidence of linear scaling of the minimal size of the training dataset required to successfully predict the outcomes of future measurements with the number of qubits in the system. 

We have tested the algorithm in a setting which is typical for most contemporary experiments with NISQ devices -- namely, for short length quantum circuits having the hardware efficient structure with alternating layers of variable single qubit gates and fixed two-qubit gates. Quantum states generated by such circuits are well described by low-rank tensor train decompositions \cite{Waintal} and our algorithm may be efficiently used to infer such decompositions from experimental data. 

All the results were obtained with a specially developed {\it QTensor} library for working with quantum states in the tensor-train format using the PyTorch framework. The package is available on the GitHub \cite{QTensor}.

While preparing the manuscript, we became aware of a closely related work \cite{li2023efficient} that relies on a similar method, but focuses on the quantum state reconstruction rather than predicting the measurement outcome probability. This fact emphasizes the keen interest for this topic in the scientific community.

\section{\label{sec:Acknowledgements}Acknowledgements}
The authors acknowledge support by Rosatom in the framework of the Roadmap for Quantum computing (Contract No. 868-1.3-15/15-2021 dated October 5, 2021 and Contract No.P2154 dated November 24, 2021). We are grateful to G.I.~Struchalin, M.Yu~Saygin and S.A.~Fldzhyan for enlightening discussions.

\bibliography{apssamp}

\appendix

\section{$R^2$ constraint for learning}\label{AppendixR2Constraint}

\begin{figure}[h]
\includegraphics[width=1.0\linewidth]{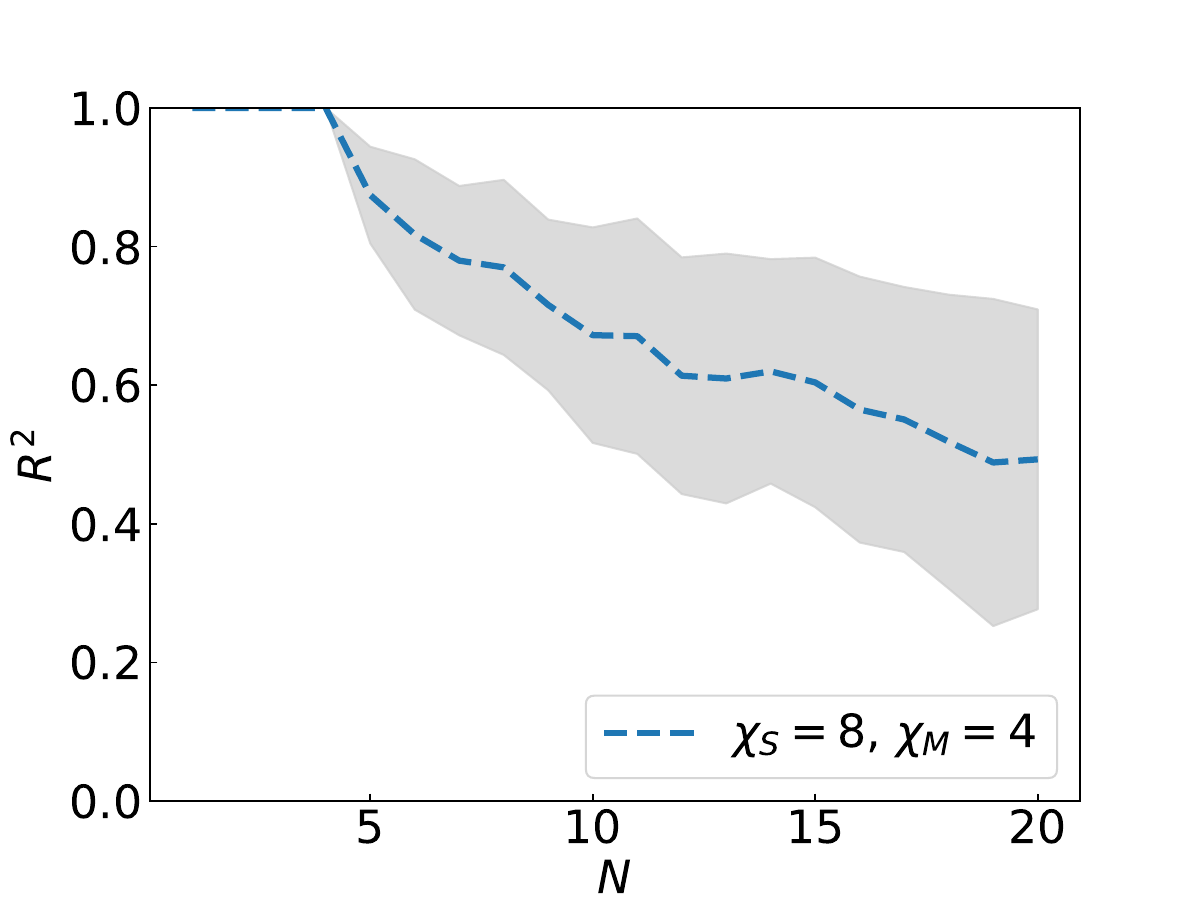}
\caption{\label{fig:MPOTruncateTest} The coefficient of determination $R^2$ averaged over $100$ different reference states with rank $\chi_S$ between the original reference states and their models with a rank limited to $\chi_M$.}
\end{figure}

When we try to train a state with a certain rank $\chi_S$ using a model with a lower rank $\chi_M$ ($\chi_M < \chi_S$), then in this case there is a certain boundary $R^2$, up to which the algorithm optimization can converge in principle. To determine this boundary $R^2$, we take random states with rank $\chi_S$ and truncate their ranks to $\chi_M$, after that we look at the resulting coefficient of determination $R^2$ between a randomly generated exact state and its truncated approximation. We calculated the coefficient of determination for $100$ $\hat{E}_i$ and averaged it over $100$ different reference states with rank $\chi_S$. The results are shown in Fig.~\ref{fig:MPOTruncateTest}.

\section{Full map $R^2$ depends on $N$ and $M$}\label{AppendixFullMapR2}

\begin{figure*}
\begin{minipage}{0.3\linewidth}
\center{\includegraphics[width=1.2\linewidth]{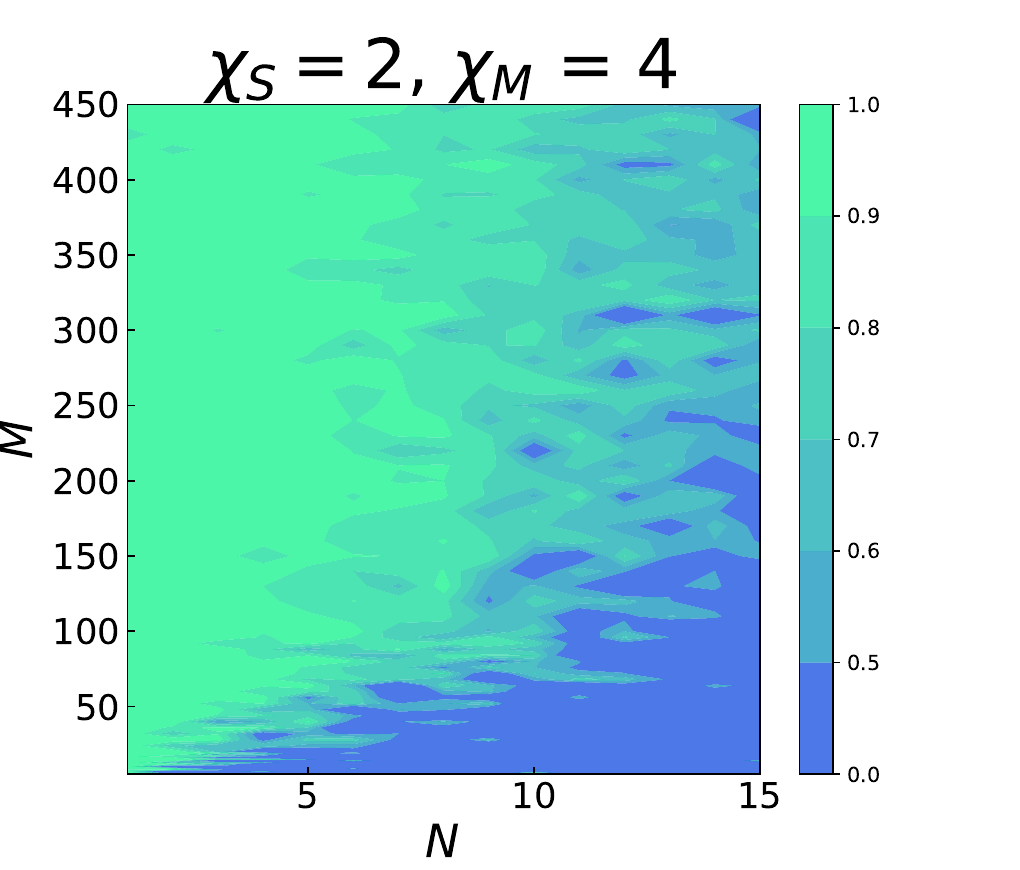}}
\end{minipage}
\begin{minipage}{0.3\linewidth}
\center{\includegraphics[width=1.2\linewidth]{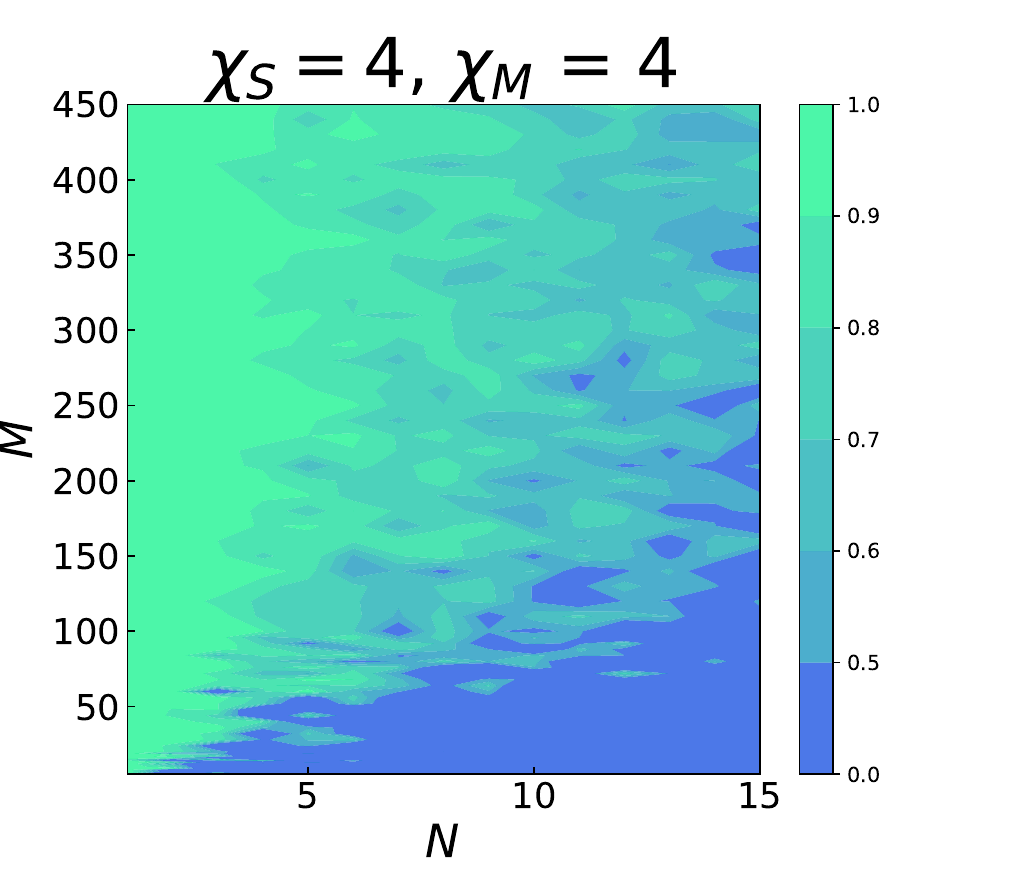}}
\end{minipage}
\begin{minipage}{0.3\linewidth}
\center{\includegraphics[width=1.2\linewidth]{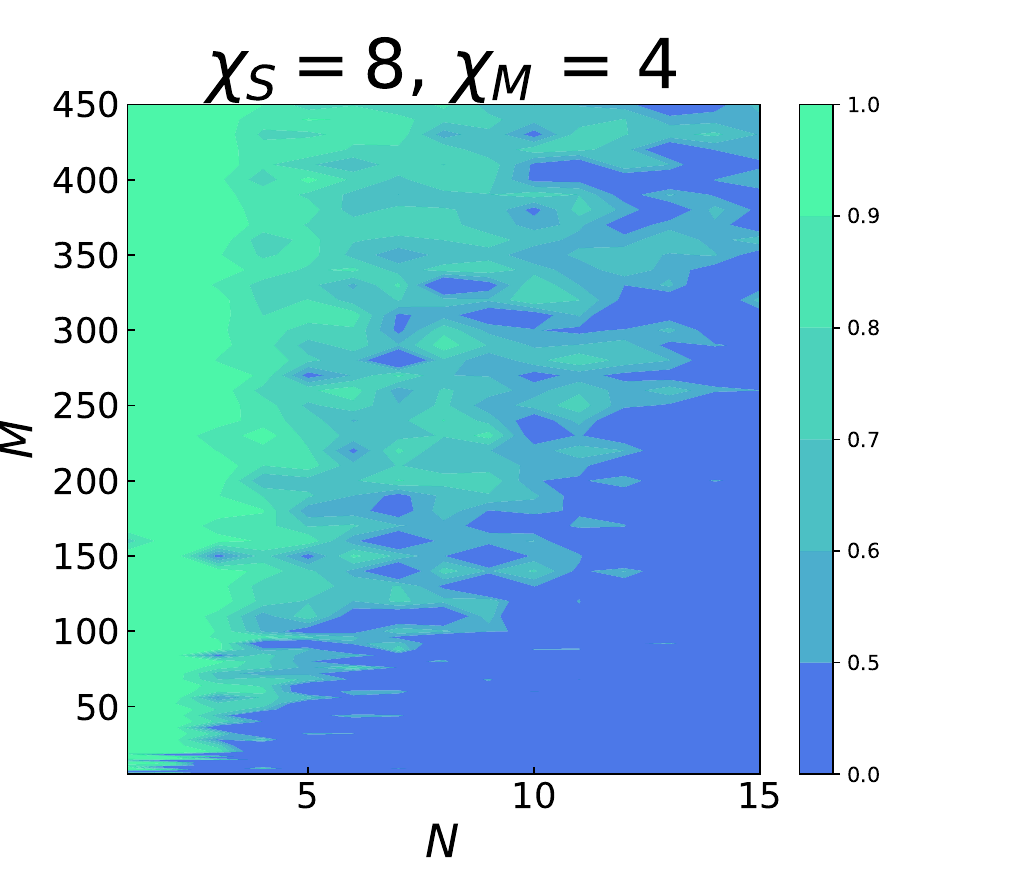}}
\end{minipage}
\caption{\label{fig:MapR2}Color map of the determination coefficient $R^2$ on the test set versus the number of qubits $N$ and the size of the training set $M$.}
\end{figure*}

In addition to Fig.~\ref{fig:NumResLinearM} in Fig.~\ref{fig:MapR2} we show a complete color map where the color displays the level of the coefficient of determination $R^2$ on the test sample. The number of qubits $N$ and the size of the training sample $M$ are plotted along the axes.

\section{Investigation of Probability Errors}\label{AppendixProbabilityErrors}

\begin{figure*}[!t]
\begin{minipage}{0.3\linewidth}
\center{\includegraphics[width=1.09\linewidth]{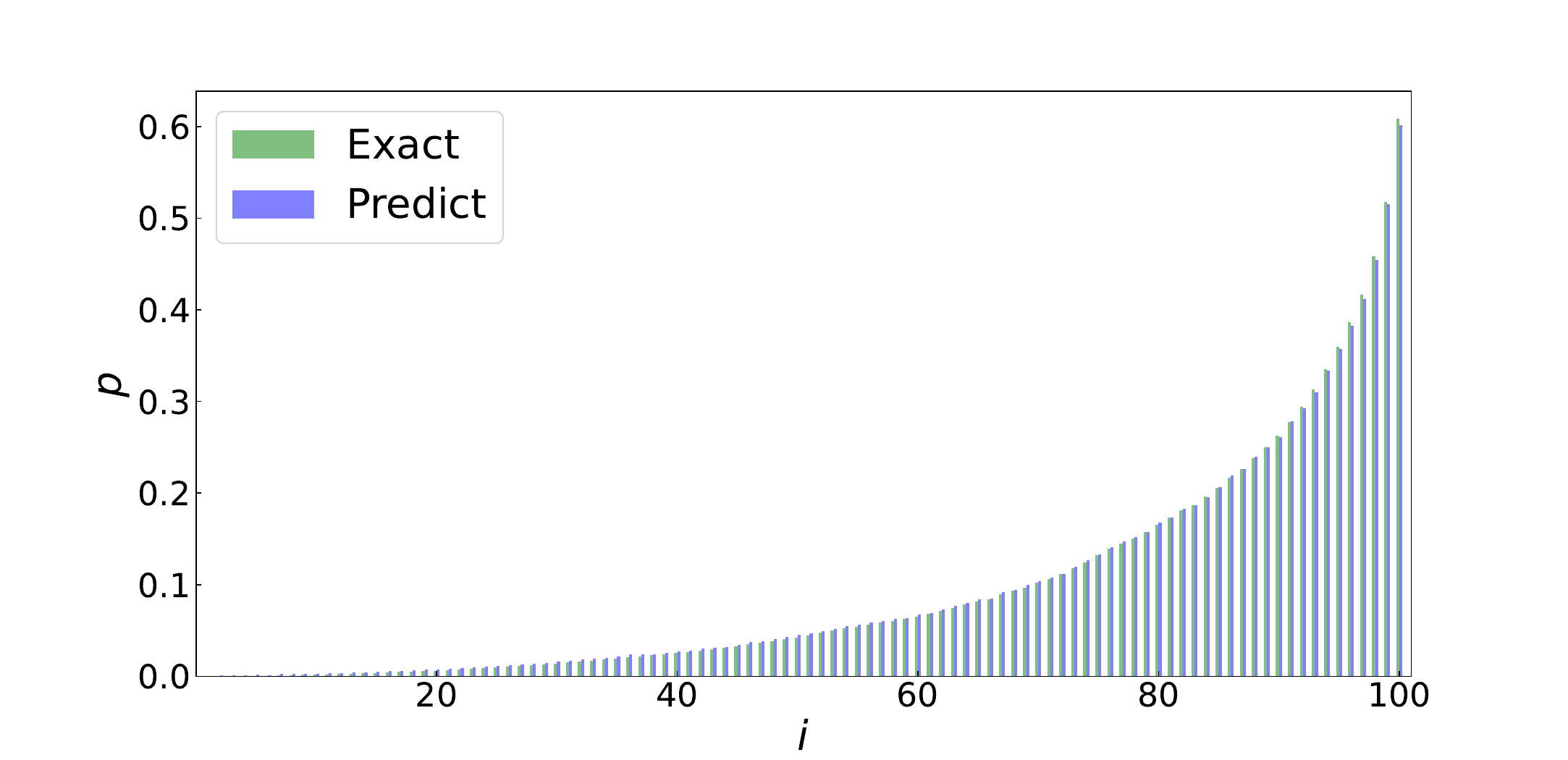}} a) Probability $\gamma = 0,0$
\end{minipage}
\begin{minipage}{0.3\linewidth}
\center{\includegraphics[width=1.09\linewidth]{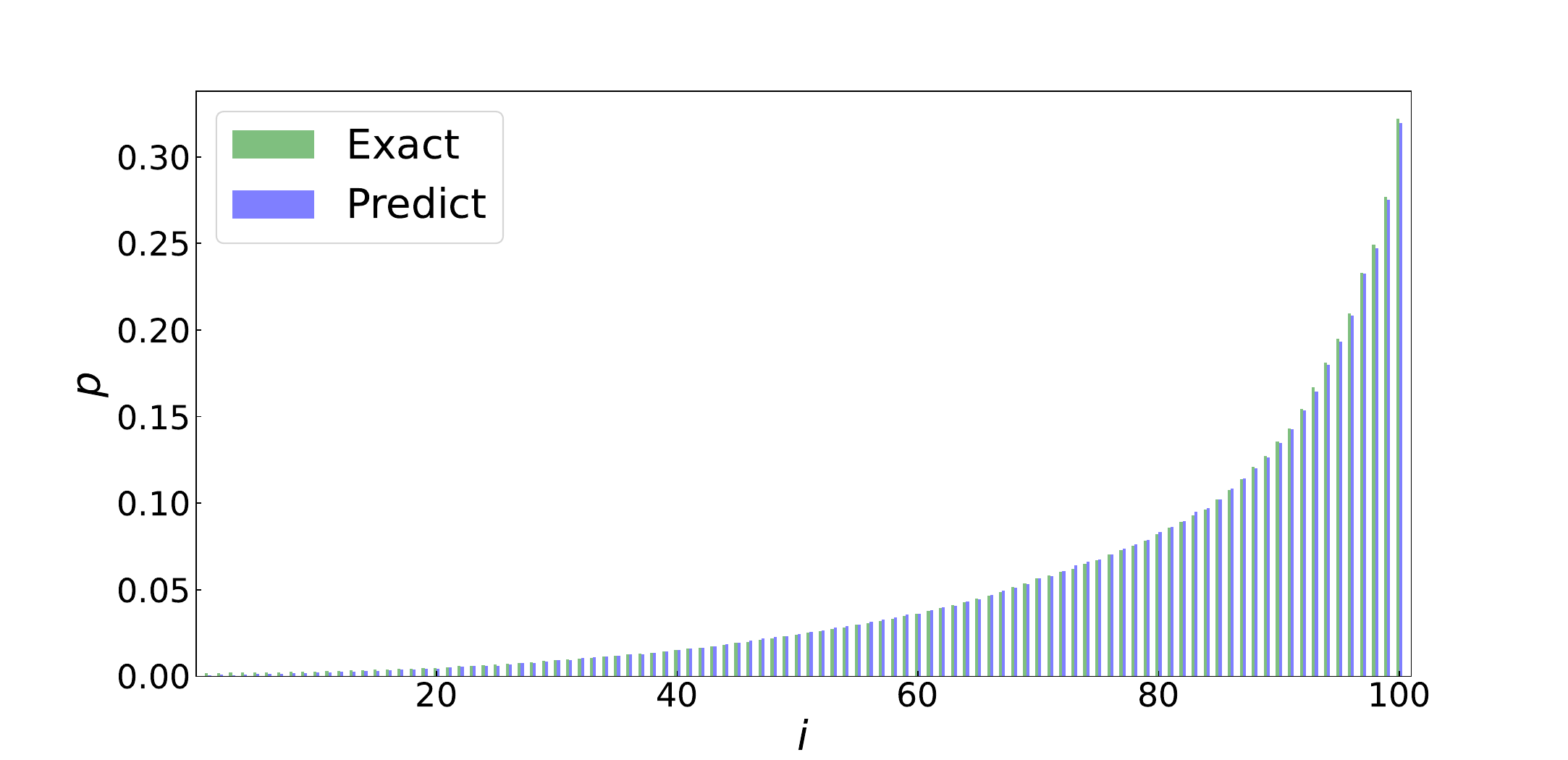}} b) Probability $\gamma = 0,5$
\end{minipage}
\begin{minipage}{0.3\linewidth}
\center{\includegraphics[width=1.09\linewidth]{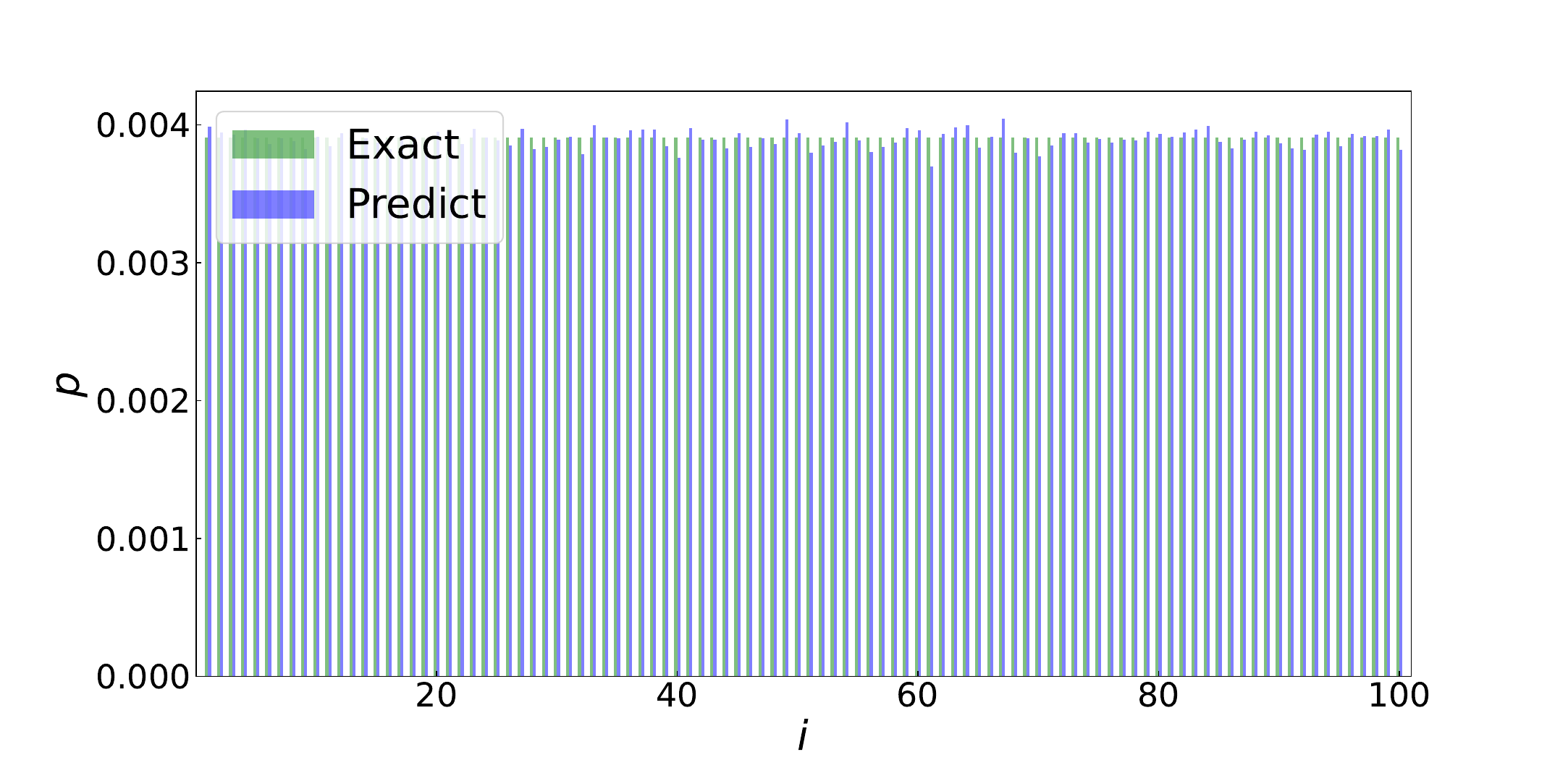}} c) Probability $\gamma = 1,0$
\end{minipage}
\begin{minipage}{0.3\linewidth}
\center{\includegraphics[width=1.09\linewidth]{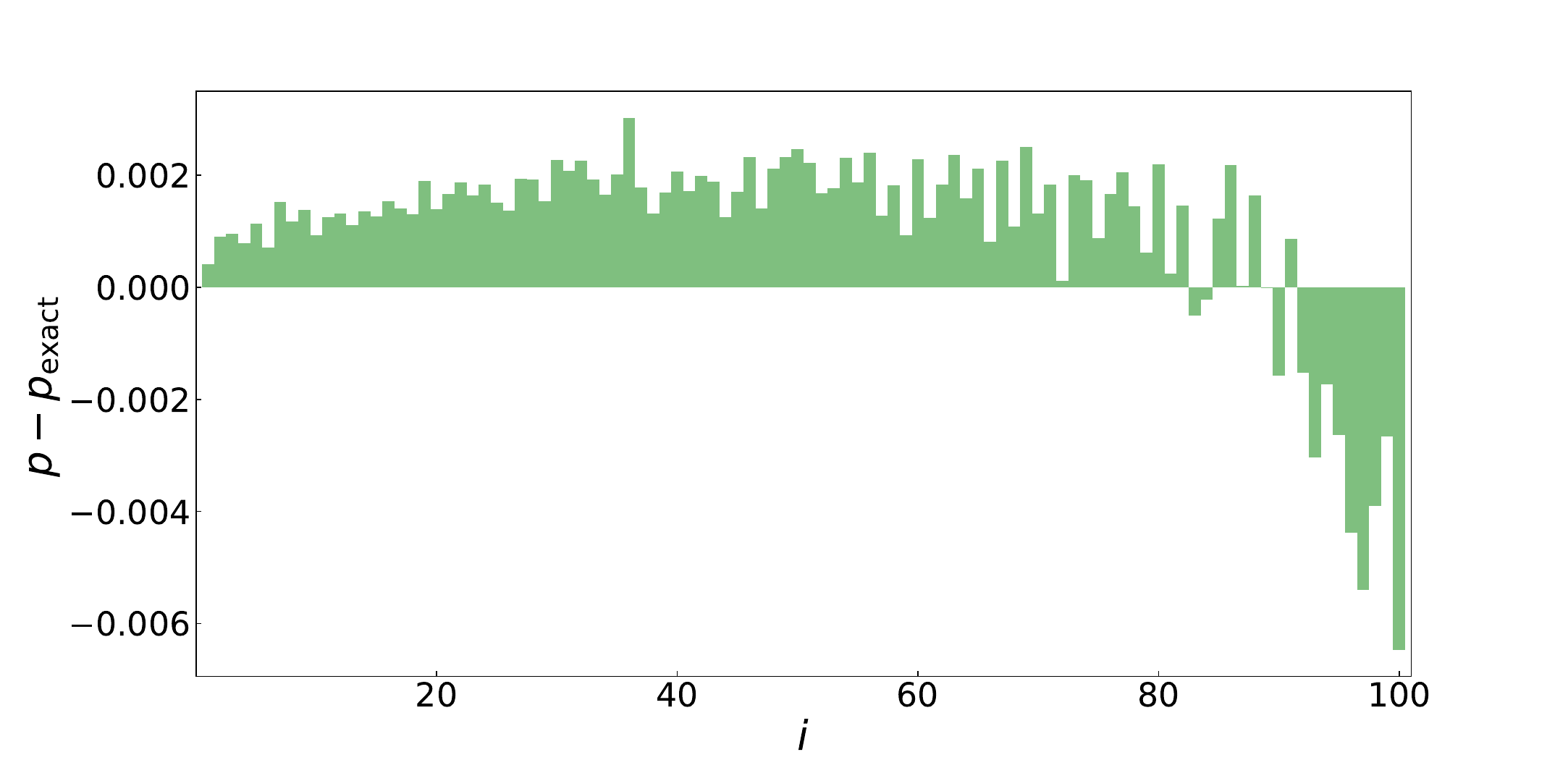}} d) Probability error $\gamma = 0,0$
\end{minipage}
\begin{minipage}{0.3\linewidth}
\center{\includegraphics[width=1.09\linewidth]{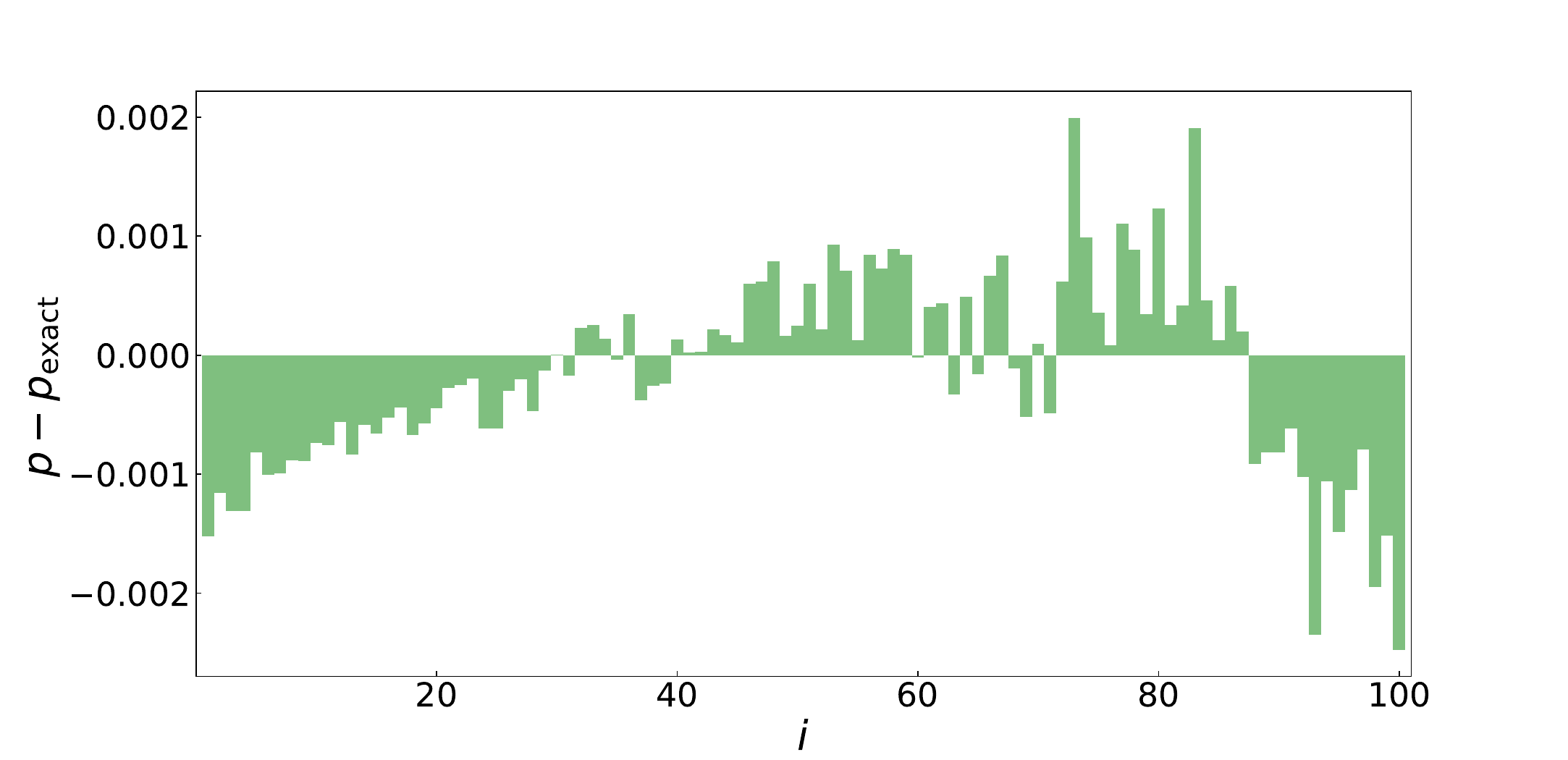}} e) Probability error $\gamma = 0,5$
\end{minipage}
\begin{minipage}{0.3\linewidth}
\center{\includegraphics[width=1.09\linewidth]{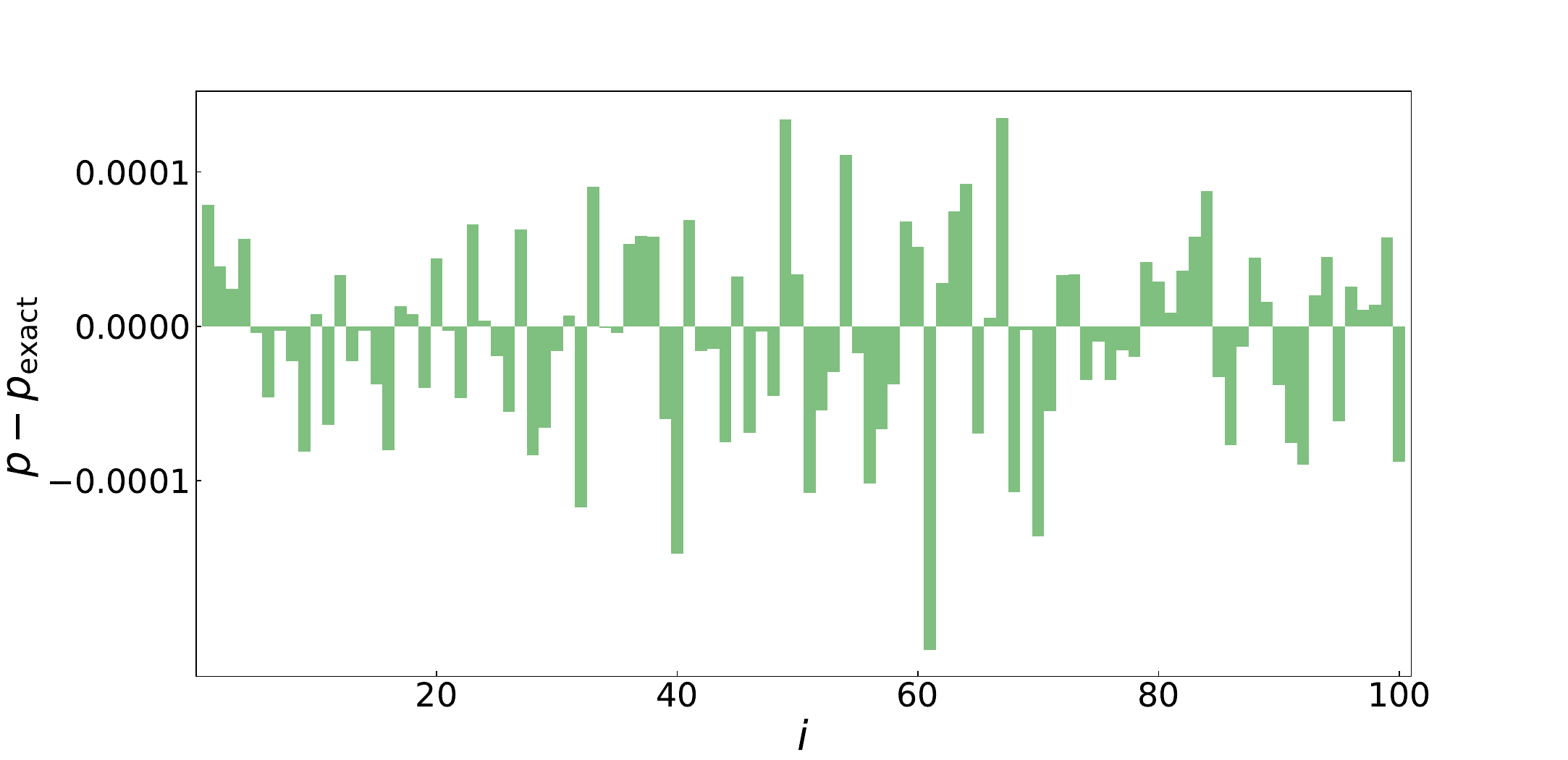}} f) Probability error $\gamma = 1,0$
\end{minipage}
\begin{minipage}{0.3\linewidth}
\center{\includegraphics[width=1.09\linewidth]{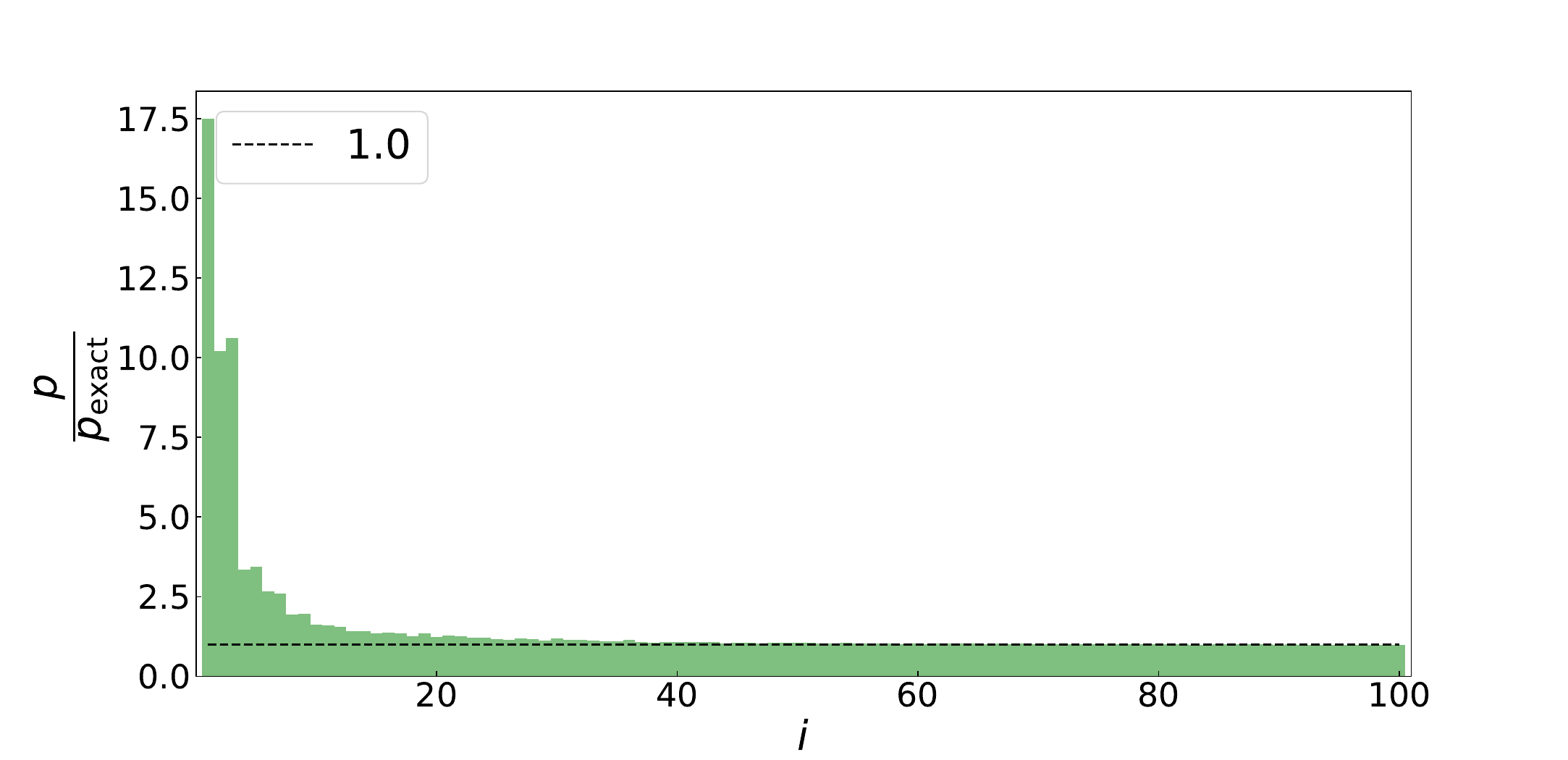}} g) Probability ratio $\gamma = 0,0$
\end{minipage}
\begin{minipage}{0.3\linewidth}
\center{\includegraphics[width=1.09\linewidth]{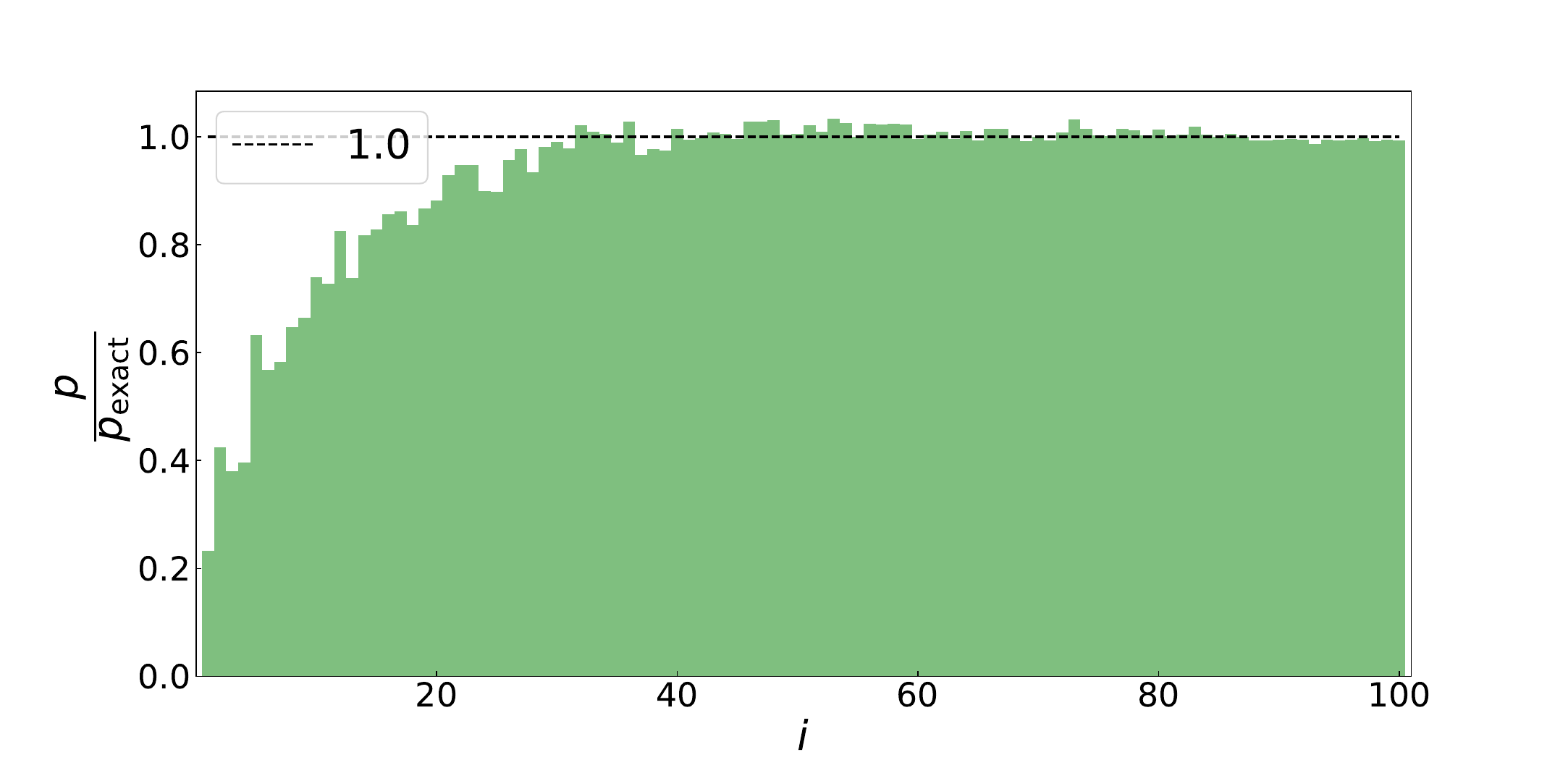}} h) Probability ratio $\gamma = 0,0$
\end{minipage}
\begin{minipage}{0.3\linewidth}
\center{\includegraphics[width=1.09\linewidth]{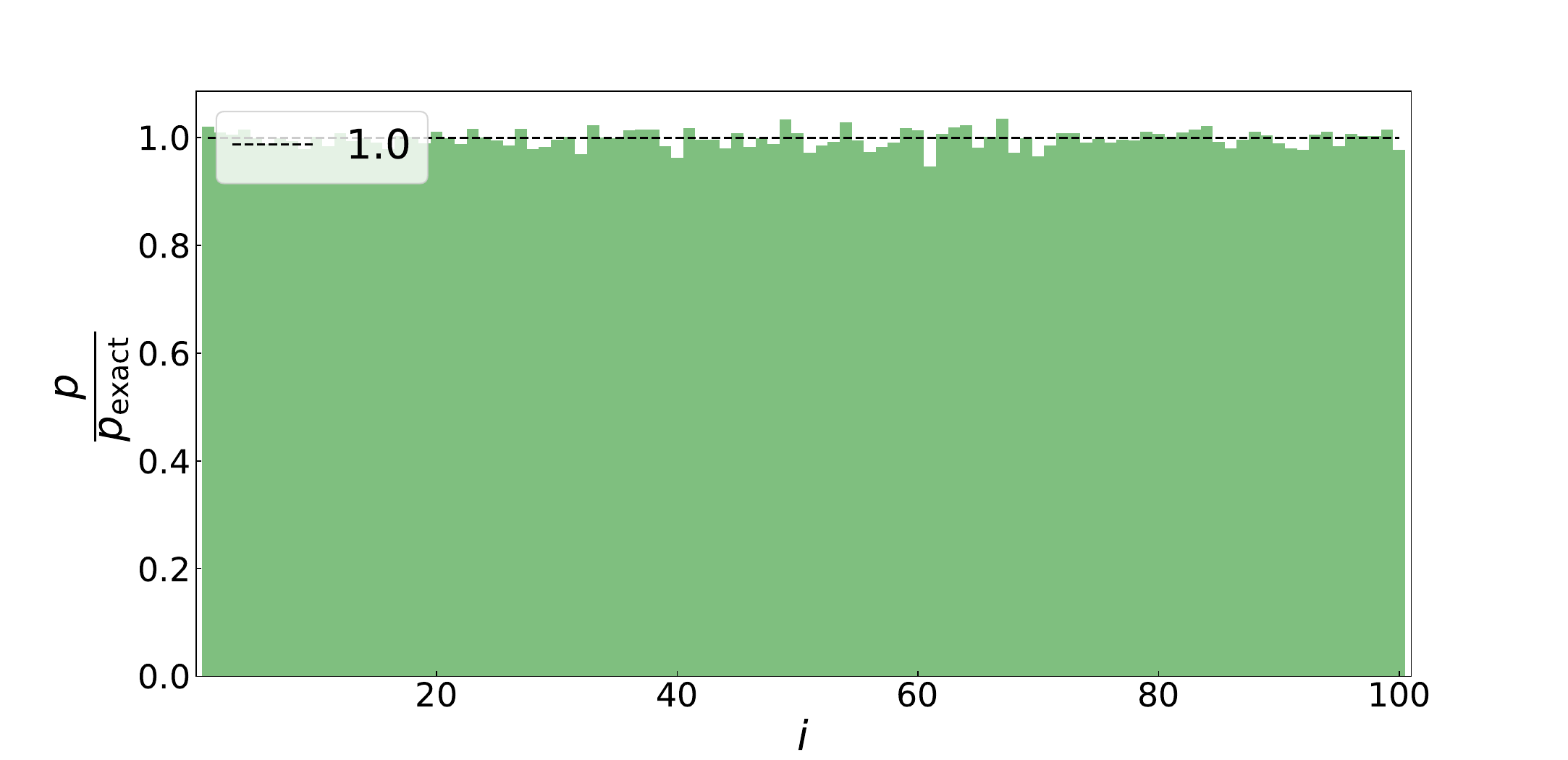}} i) Probability ratio $\gamma = 0,0$
\end{minipage}
\caption{\label{fig:ProbabilityError}Extension Fig.~\ref{fig:NumResPhysDevice}. The top plots (a, b, c) show a comparison of the exact probability distributions with the probability distribution that the model will produce. The middle graphs (d, e, f) show an error in determining the probabilities $p - p_{\operatorname{exact}}$. The lower plots (g, h, i) show the ratio of the probability that the model produces to the exact probability $\dfrac{p}{p_{\operatorname{exact}}}$. The results are shown with averaging over $30$ different sets of operators POVM $\hat{E}_i$.}
\end{figure*}

In addition to Fig.~\ref{fig:NumResPhysDevice} we study how the errors in probability predicted by our model are distributed depending on the magnitude of the probability. Differences and the ratios of probabilities are shown in Fig.~\ref{fig:ProbabilityError}.

\begin{figure*}
\begin{minipage}{0.3\linewidth}
\center{\includegraphics[width=1.09\linewidth]{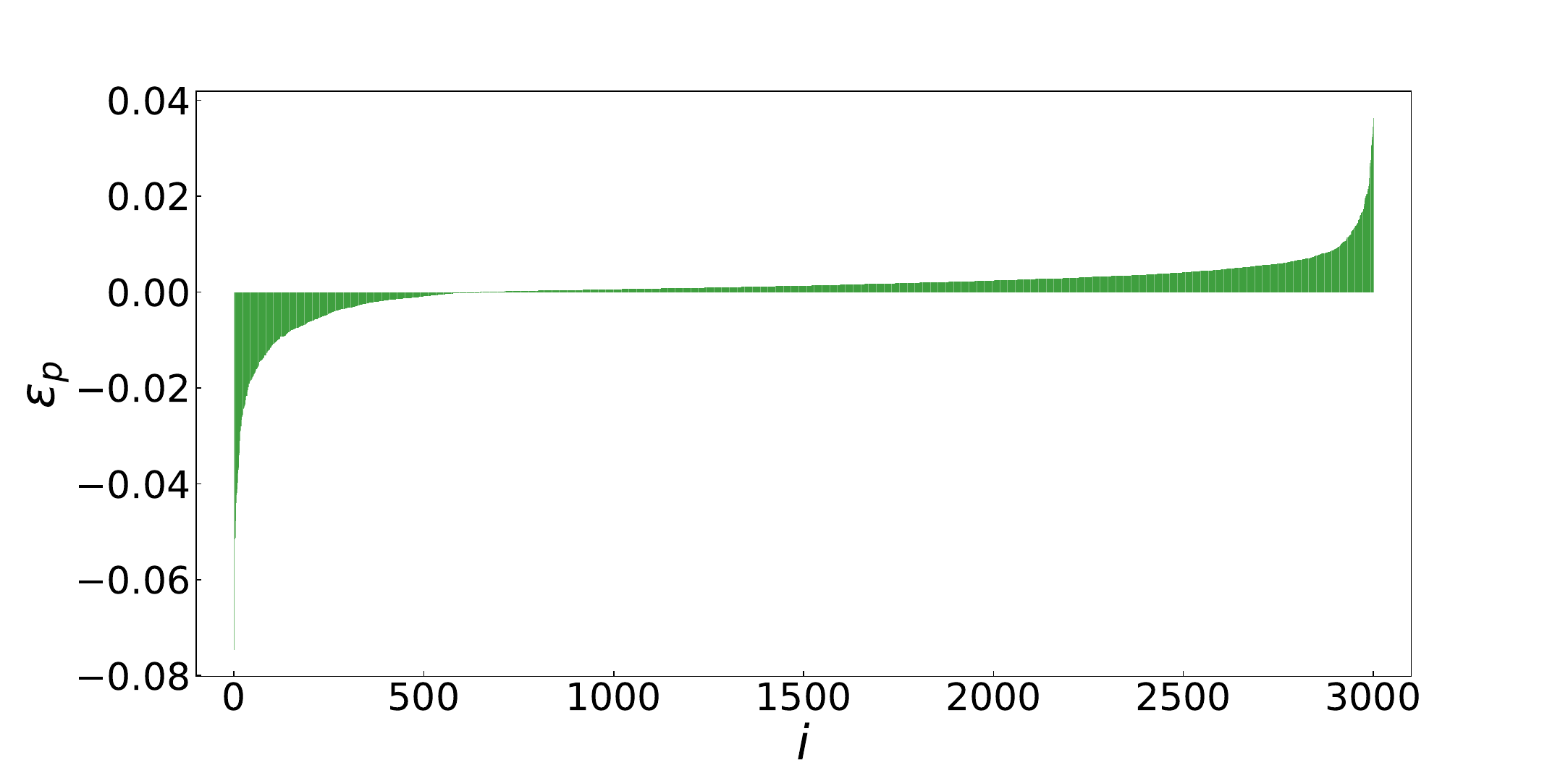}} a) $\gamma = 0,0$
\end{minipage}
\begin{minipage}{0.3\linewidth}
\center{\includegraphics[width=1.09\linewidth]{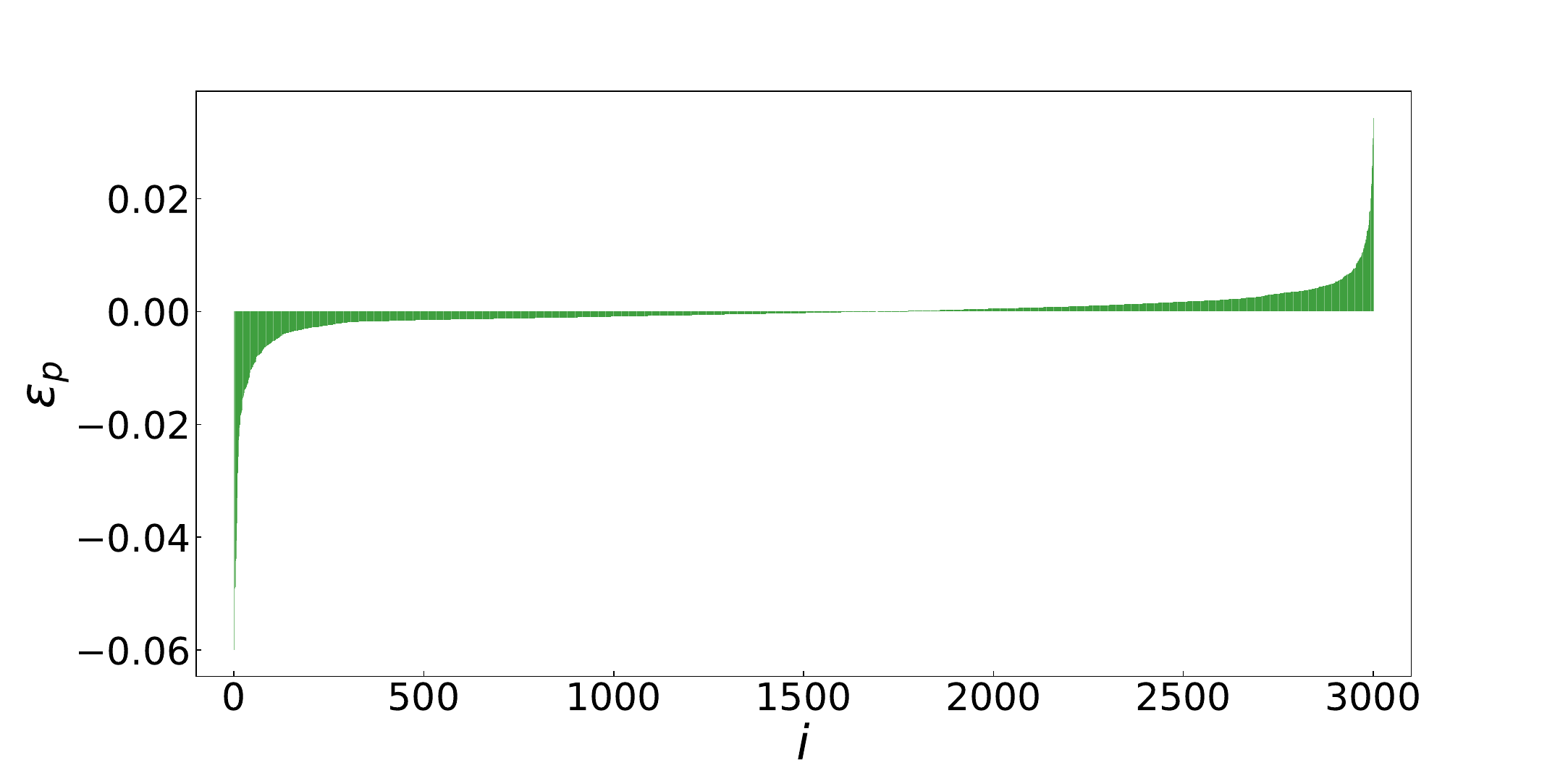}} b) $\gamma = 0,5$
\end{minipage}
\begin{minipage}{0.3\linewidth}
\center{\includegraphics[width=1.09\linewidth]{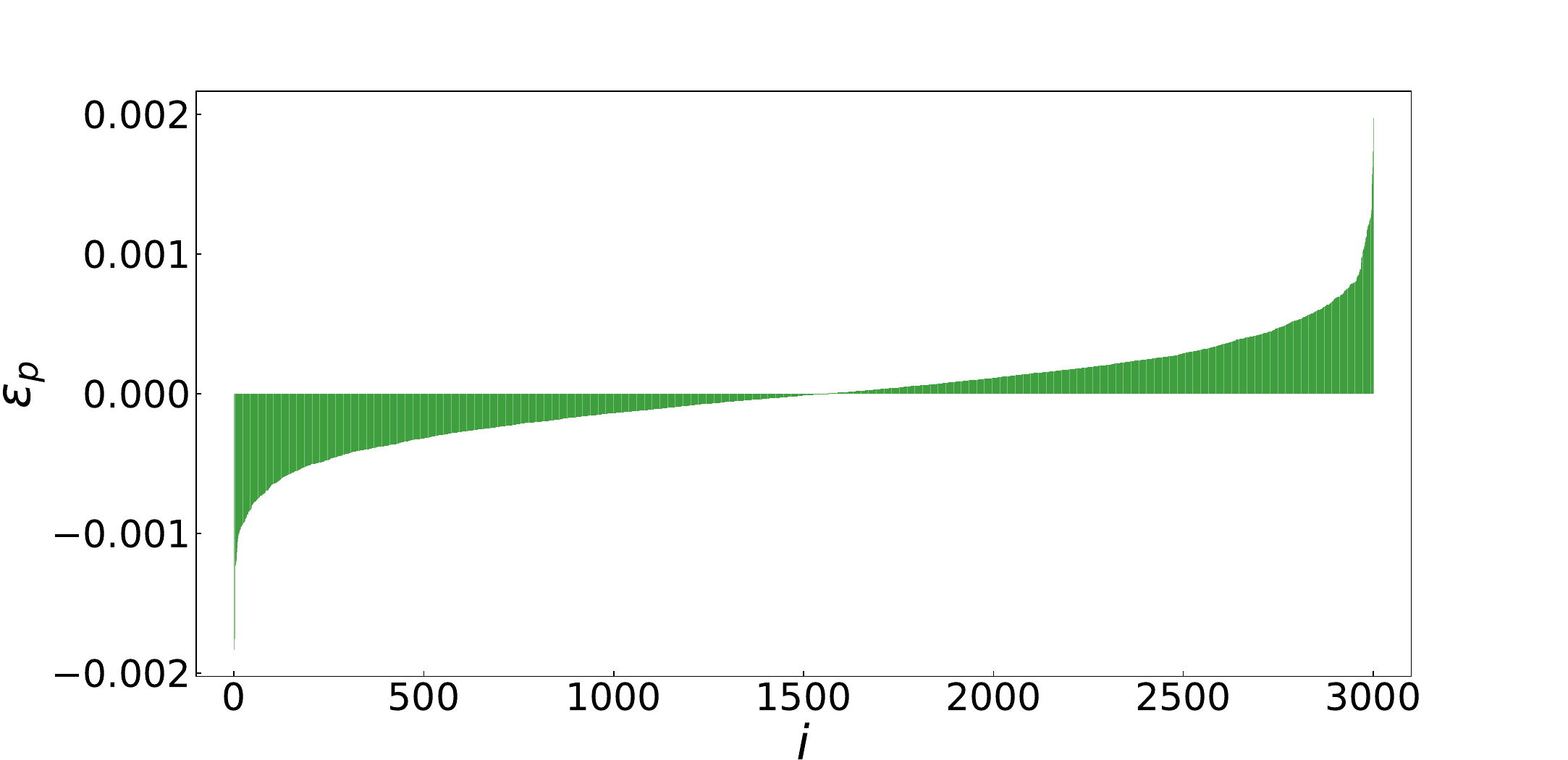}} c) $\gamma = 1,0$
\end{minipage}
\caption{\label{fig:ProbErrorDistr}Probability error distribution function $\varepsilon_p = p_{\operatorname{predict}} - p_{\operatorname{exact}}$, obtained by considering $30$ different states generated according to the scheme in Fig.~\ref{Depolarization_channel} and measured over $100$ different POVM operators. To estimate the share of strong outliers, we calculated for each case (a, b, c) the sample standard deviation $\sigma_{\varepsilon_p}$ and calculated the proportion of cases where $|\varepsilon_p| > 3\sigma_{\varepsilon_p}$: a) $2,2\%$, b) $1,9\%$, c) $1,0\%$.}

\end{figure*}

Also of interest is the question of how our model fits within the PAC framework. To answer this question, we calculated the distribution of the errors in predicted probabilities. It is shown in Fig.~\ref{fig:ProbErrorDistr}. At the edges of this distribution, we observe narrow tails down and up, which indicates that the trained model with some small probability can be very wrong in predicting the probability. This failure mode of our model is in line with general expectations for PAC learning.

\nocite{*}


\end{document}